\documentclass[aps,prx,reprint,superscriptaddress]{revtex4-2}
\usepackage{blindtext}
\usepackage{graphicx}
\usepackage{amsmath}
\usepackage{amssymb}
\usepackage[english]{babel}
\usepackage{xcolor}
\usepackage{siunitx}
\usepackage[colorlinks=true, pdfstartview=FitV, linkcolor=blue,
citecolor=blue, urlcolor=blue]{hyperref}
\usepackage[capitalise]{cleveref}

\usepackage{float}

\makeatletter
\let\saved@includegraphics\includegraphics
\AtBeginDocument{\let\includegraphics\saved@includegraphics}
\renewenvironment*{figure}{\@float{figure}}{\end@float}

\makeatother

\makeatletter
\newcommand\footnoteref[1]{\protected@xdef\@thefnmark{\ref{#1}}\@footnotemark}
\makeatother	

\begin{document}

\title{Fully tunable longitudinal spin-photon interactions in Si and Ge quantum dots}

\author{Stefano Bosco}
\email{stefano.bosco@unibas.ch}
\affiliation{Department of Physics, University of Basel, Klingelbergstrasse 82, 4056 Basel, Switzerland}

\author{Pasquale Scarlino}
\affiliation{Institute of Physics, Ecole Polytechnique F\'ed\'erale de Lausanne, CH-1015 Lausanne, Switzerland}
\author{Jelena Klinovaja}
\affiliation{Department of Physics, University of Basel, Klingelbergstrasse 82, 4056 Basel, Switzerland}
\author{Daniel Loss}
\affiliation{Department of Physics, University of Basel, Klingelbergstrasse 82, 4056 Basel, Switzerland}

\begin{abstract}
Spin qubits in silicon and germanium quantum dots are promising platforms for quantum computing, but  entangling spin qubits over micrometer distances remains a critical challenge. 
Current prototypical architectures  maximize transversal interactions between qubits and  microwave resonators, where the spin state is flipped by nearly resonant photons. However, these interactions cause back-action on the qubit, that yield unavoidable residual qubit-qubit couplings and significantly affect the gate fidelity.
Strikingly, residual couplings vanish when spin-photon interactions are longitudinal and  photons couple to the phase of the qubit.
We show that large longitudinal interactions emerge naturally in state-of-the-art hole spin qubits. These interactions are fully tunable and can be parametrically modulated by external oscillating electric fields. 
We propose realistic protocols to measure these interactions and to implement fast and high-fidelity two-qubit entangling gates. These protocols work also at high temperatures,  paving the way towards the implementation of  large-scale quantum processors.
\end{abstract}

\maketitle

\paragraph*{Introduction.}

Spin qubits in silicon (Si) and germanium (Ge) quantum dots are frontrunner candidates to process quantum information~\cite{scappucci2020germanium,doi:10.1146/annurev-conmatphys-030212-184248,burkard2021semiconductor,zwerver2021qubits,mills2021two,Jirovec2021}.
Hole spin qubits hold particular promise because of their large and fully tunable spin-orbit interactions (SOI)~\cite{PhysRevLett.98.097202,PhysRevLett.95.076805,DRkloeffel1,DRkloeffel3,PRXQuantum.2.010348,PhysRevB.99.115317,PhysRevB.104.115425,PhysRevB.103.125201,Wang2021}, enabling ultrafast all-electrical gates at low power~\cite{watzinger2018germanium,  Hendrickxsingleholespinqubit2019,hendrickx2020fast,hendrickx2020four,Wang2022,Froning2021,camenzind2021spin,maurand2016cmos,voisin2016electrical,piot2022single},
and because of their resilience to  hyperfine noise even in natural materials~\cite{PhysRevLett.127.190501,PhysRevLett.105.266603,PhysRevB.78.155329,prechtel2016decoupling,warburton2013single,PhysRevB.101.115302}.
In current quantum processors, engineering long-range interactions of distant qubits remains a critical challenge.  
A fast and coherent interface between qubits separated by a few micrometers will enable modular architectures with cryogenic classical control on-chip~\cite{VandersypenInterfacingspinqubits2017,Gonzalez-Zalba2021,Xue2021,8993570,boter2021spider}, as well as significantly improve qubit connectivity, with great advantages for near-term quantum processors~\cite{doi:10.1073/pnas.1618020114,Holmes_2020} and opening up to new classes of efficient quantum error correcting codes~\cite{PRXQuantum.2.040101,10.5555/2685179.2685184,cohen2021low}.

Driven by significant technological progress in enhancing the amplitude of spin-photon interactions~\cite{Landig2018,mi2018coherent,doi:10.1126/science.aaa3786}, coupling distant spin qubits via microwave resonators is an appealing approach~\cite{harvey2021circuit}. 
In analogy to superconducting circuits~\cite{RevModPhys.93.025005,PhysRevA.69.062320}, current spin-photon interfaces are designed to be \textit{transversal}, where nearly resonant photons flip the spin state. However, these interactions cause a significant back-action of the resonator on the qubit, resulting in unavoidable residual qubit-qubit couplings. These unwanted couplings are critical issues in scalable quantum processors~\cite{PhysRevB.91.094517,PhysRevB.93.134501}, and they are minimized by operating in the dispersive regime~\cite{PhysRevA.79.013819}, where a large detuning between resonator and qubit frequencies suppresses also the effective spin-spin interactions.

In this work, we investigate a different approach where fast high-fidelity two-qubit  gates are implemented by tunable \textit{longitudinal} spin-photon interactions, where the photon couples to the phase of the qubit.
These interactions do not cause back-action on the qubit, thus eliminating residual couplings and suppressing Purcell decay~\cite{PhysRevB.91.094517}. Moreover, because longitudinal interactions do not rely on resonant processes, fast two-qubit gates are possible at arbitrarily large detuning, relaxing stringent constraints imposed by dispersive interactions and frequency crowding, and helping in scaling up the next-generation of quantum processors.

While challenging to engineer in superconducting circuits~\cite{PhysRevB.91.094517,PhysRevB.93.134501}, large longitudinal interactions emerge naturally in hole spin qubits, where, in  contrast to alternative theoretical proposals~\cite{PhysRevLett.108.190506,bottcher2021parametric,PhysRevB.97.235409,Beaudoin_2016,PhysRevB.100.035416,PhysRevB.103.035301} and recent experiments~\cite{bottcher2021parametric}, they  do not require multiple quantum dots, nor parametric driving.
 We show that longitudinal interactions in hole  dots are fully tunable and can be modulated by oscillating electric fields, enabling significantly faster qubit readout protocols~\cite{PhysRevLett.115.203601,PhysRevB.99.245306} and  two-qubit gates~\cite{Royer2017fasthighfidelity,PhysRevB.96.115407}.
We propose protocols to reliably measure and control these interactions, yielding fast entangling gates between distant qubits, with high fidelities above the surface code threshold~\cite{PhysRevLett.109.180502}. Strikingly, these gates work also at high temperatures with resonators in hot thermal states~\cite{PhysRevLett.126.250505}, thus providing a significant step towards large-scale universal semiconducting quantum processors.

\paragraph*{Static spin-photon interactions.}

\begin{figure}
\centering
\includegraphics[width=0.5\textwidth]{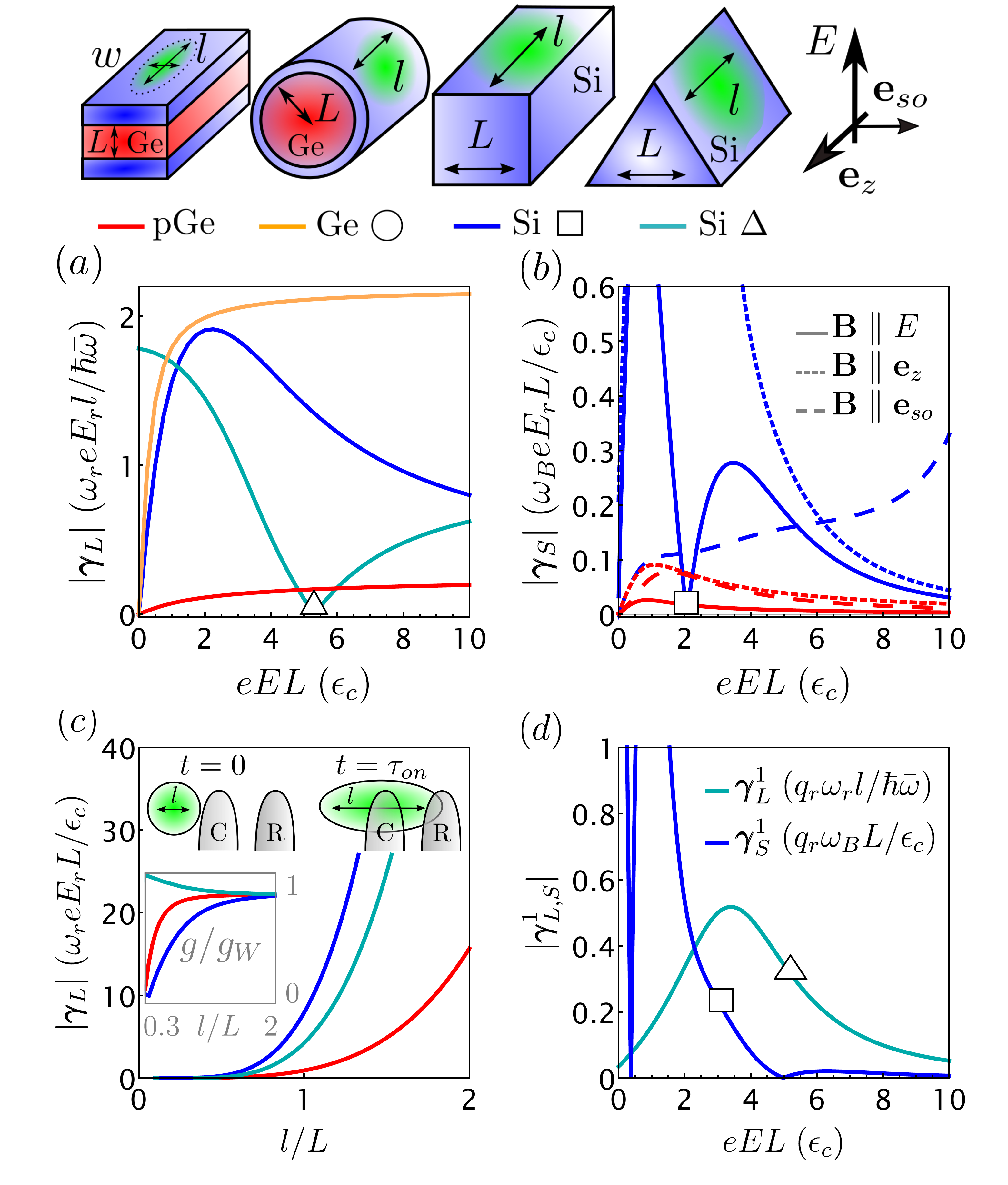}
\caption{\label{fig:static-coupling} 
Interactions of hole spin qubits and microwave photons. At the top, we sketch the devices analyzed and the color code used.
 In (a) and (b) we show the magnitude of the static interactions $\pmb{\gamma}_L$ and $\pmb{\gamma}_S$ against $E$.
In state-of-the-art architectures, $\pmb{\gamma}\sim 100$~MHz, and $E$ is in the V/$\mu$m range, enabling strong spin-photon interactions. 
In~(c) and~(d), we examine the tunability of $\pmb{\gamma}$. In~(c), we show how to turn on/off $\pmb{\gamma}_L$ by modulating  $l$  in a time  $\tau_{on}$. A control gate C tunes $l$ and, in the off state,  screens the gate R connected to the resonator. In the inset, we show the variation of the $g$-factor during the protocol. In Si and Ge $eEL=3\epsilon_c$ and $eEL=5\epsilon_c$, respectively, and $g_W^\text{pGe}=0.27$, $g_W^{\text{Si}\square}=1.15$, and $g_W^{\text{Si}\Delta}=2.8$.
In (d), we show the ac modulations of the interactions $\pmb{\gamma}^1_{L,S}$.
Squares and triangles  indicate  where $\pmb{\gamma}^0=0$.
Here, $q_r=e^2E_r L/\epsilon_c$ and $\epsilon_c=\hbar^2\pi^2/\bar{m}L^2$.
 The Ge qubits are encoded in  squeezed  dots~\cite{PhysRevB.104.1154251} in planar Ge/SiGe heterostructures with $L=30$~nm,  $l=5w=50$~nm, and strain energy $\epsilon_s=15$~meV~\cite{lodari2021lightly}, and in Ge/Si core/shell nanowires with $l=5L=50$~nm and $\epsilon_s=25$~meV~\cite{adelsberger2021hole}. The Si qubits are encoded in square and  triangular finFETs with  $l=2L=20$~nm. We consider isotropic Ge~\cite{WinklerSpinOrbitCoupling2003}, and in Si we use the growth direction $\textbf{e}_z\parallel [001]$, $E\parallel [110]$, where  SOIs are maximal~\cite{DRkloeffel3} (fully tunable~\cite{PRXQuantum.2.010348, PhysRevLett.127.190501}) in square (triangular)~FETs.
}
\end{figure}

The interactions between spins in a quantum dot and  photons confined in a microwave resonator with frequency $\omega_r$ is described by the Hamiltonian 
\begin{equation}
\label{eq:Ham}
H=\frac{ \hbar \omega_B}{2}\sigma_3+\hbar\omega_r a^\dagger a+\hbar\pmb{\gamma}\cdot\pmb{\sigma}(a^\dagger+a) \ ,
\end{equation}
where $a$ is the photon annihilation operator and $\pmb{\sigma}$ is a vector of Pauli matrices. The qubit frequency $\omega_B$ and the direction of Zeeman field $\textbf{e}_3$ are related to the magnetic field $\textbf{B}$ and to the matrix $\underline{g}$ of quantum dot $g$-factors  by $\mu_B\textbf{B}\cdot\underline{g}=\hbar \omega_B \textbf{e}_3$.
The spin-photon interactions are determined by the vector $\pmb{\gamma}$, whose magnitude and direction strongly depends on the setup.

In long quantum dots, where the confinement potential in one direction, e.g. $\textbf{e}_z$, is  smoother than in the other directions, the spin-photon interactions are mediated by the SOI, and when the electric field $E_r$ of the resonator is aligned to $\textbf{e}_z$, we obtain~\cite{SM}
\begin{equation}
\label{eq:long-dot-coupling}
\pmb{\gamma}_L=\frac{l}{l_{so}}\frac{eE_r l}{\hbar\bar{\omega}} \omega_r\textbf{e}_{so} \approx \alpha\frac{l}{l_{so}}\frac{\sqrt{Z_r}}{\sqrt{\hbar}\bar{\omega}}\omega_r^2 \textbf{e}_{so} \ .
\end{equation}
The spin-orbit length $l_{so}$ and the unit vector $\textbf{e}_{so}$ describe the magnitude and direction of the vector of SOI,  respectively. We parametrize the soft and hard confinement by the harmonic length $l$ and frequency $\bar{\omega}=\hbar/\bar{m}l^2$, and by the length $L$ and energy  $\epsilon_c=\hbar^2\pi^2/\bar{m}L^2$, respectively. Also, in hole quantum dots the average mass is $\bar{m}=m/\gamma_1$, where $m$ is the bare electron mass and $\gamma_1$ is a Luttinger parameter~\cite{WinklerSpinOrbitCoupling2003}.
We estimate $e E_r l\approx \alpha V_r$, where $V_r=\omega_r\sqrt{\hbar Z_r}$ is the zero-point-fluctuation potential of  photons at the antinode of a resonator with characteristic impedance $Z_r$~\cite{PhysRevA.69.062320,RevModPhys.93.025005} and $\alpha$ is the lever arm describing the electrostatic coupling of the plunger gate connecting the resonator to the dot~\footnote{We consider an electrode R at distance $\lambda$ from the dot and connected to the resonator. By introducing $\alpha=e C_R/C_{D}\approx e l /\lambda $, with gate and dot capacitances $C_R\sim 1/\lambda$ and  $C_D\sim 1/l$, we estimate  $E_r\approx V_r/\lambda\approx \alpha V_r/e l $. }. 

The type of spin-photon interactions depends on the direction of $\textbf{B}$. Transversal interactions are tuned into longitudinal interactions by  aligning the Zeeman field to the spin-orbit vector, i.e. $\textbf{e}_3\parallel\textbf{e}_{so}$.
The magnitude of $\pmb{\gamma}_L$ is particularly large in hole spin qubits, where $l_{so}$ of a few tens of nanometers, comparable to $l$, were measured~\cite{Froning2021,Wang2022,PhysRevResearch.3.013081}, enabling strong interactions  in single quantum dots~\cite{DRkloeffel2,bosco_loss2022,PhysRevB.102.205412}.

In Fig.~\ref{fig:static-coupling}(a), we compare $\pmb{\gamma}_L$ in state-of-the-art hole spin qubits encoded in Si and Ge quantum dots. More details on the simulation are given in Sec.~I of~\cite{SM}. 
Importantly, the amplitude of $\pmb{\gamma}_L$ is large and fully tunable by an external electric field $E$.
By simulating long quantum dots in Ge/Si core/shell nanowires~\cite{DRkloeffel1,DRkloeffel2,Froning2021,PhysRevResearch.3.013081} and in square Si finFETs~\cite{maurand2016cmos,voisin2016electrical,piot2022single}, we observe that large coupling strengths $|\pmb{\gamma}_L^{\text{GeO}}|/2\pi\approx 200$~MHz and $|\pmb{\gamma}_L^{\text{Si} \square}|/2\pi\approx 100$~MHz are experimentally achievable at realistic fields $E\sim 10$~V/$\mu$m and for typical parameters $\alpha=0.4e$, $\omega_r/2\pi=5$~GHz and $Z_r=4$~k$\Omega$ (yielding $V_r=20$~$\mu$V)~\cite{PhysRevApplied.5.044004}. 
At small $E$, similar values of $|\pmb{\gamma}_L|$ emerge  in triangular Si finFETs~\cite{kuhlmann2018ambipolar,geyer2020silicon,camenzind2021spin,AK_pr}, where the lower symmetry of the cross-section permits to completely turn off the SOI at finite values of $E$~\cite{PRXQuantum.2.010348} (marked with a triangle).
We also examine a quantum dot in strained planar Ge/SiGe heterostructures, an architecture that holds much promise  for scaling up quantum computers~\cite{Hendrickxsingleholespinqubit2019,hendrickx2020fast,hendrickx2020four}. Squeezing the quantum dot, such that the harmonic lengths defining the dot are $w$ and $l$, and satisfy $w\ll l$, induces a large SOI~\cite{PhysRevB.104.115425}, and enables significant spin-photon interactions, with $|\pmb{\gamma}_L^\text{pGe}|/2\pi \approx 20$~MHz. 

These interactions are enhanced by  modifying the quantum dot, e.g. using lightly strained materials and tightly squeezed dots~\cite{PhysRevB.104.115425}, or by increasing the resonators impedance, either in superconducting platforms~\cite{PhysRevApplied.11.044014,Grunhaupt2019,Maleeva2018,PhysRevLett.121.117001} or in carbon nanotubes~\cite{doi:10.1063/1.4868868,1406008,Chudow2016} and quantum Hall materials~\cite{PhysRevB.100.035416,PhysRevApplied.12.014030,PhysRevResearch.2.043383,PhysRevB.96.115407}, where $Z_r\approx 25$~k$\Omega$. Also, importantly, $\pmb{\gamma}_L$ depends quadratically on the resonator frequency, see Eq.~\eqref{eq:long-dot-coupling}, and increasing $\omega_r$ has a large effect on the interactions, enabling $|\pmb{\gamma}_L|\sim 1$ GHz at $\omega_r/2\pi=20$~GHz. 
In striking contrast to transversal interactions, longitudinal interactions do not require $\omega_r$ and $\omega_B$ to be matched~\cite{Royer2017fasthighfidelity}, and working at $\omega_r/2\pi\sim 20$~GHz and $\omega_B/2\pi\sim 1$~GHz suppresses unwanted residual dispersive couplings $\chi$, that  affect scalability~\cite{PhysRevB.91.094517,PhysRevB.93.134501}. When $\textbf{B}$ is misaligned by $\eta=1\%$ from $\textbf{e}_{so}$, we estimate $\chi\approx \eta^2 |\pmb{\gamma}_L|^2/(\omega_r-\omega_B)\lesssim 10^{-5} |\pmb{\gamma}_L|$. Moreover, in this regime, small values of $\textbf{B}$ suffice to strongly couple the spin to the resonator, yielding longer photon lifetime in superconducting cavities and strong interactions in qubit architectures with small $g$-factors~\cite{Hendrickxsingleholespinqubit2019,hendrickx2020fast,hendrickx2020four}.

In hole spin qubits, large spin-photon interactions arise also from the electrical tunability of the Zeeman energy~\cite{Froning2021}. When $\textbf{B}$ is misaligned to a principal axis $i$ of the $g$-factor (defined by $\underline{g}=\delta_{ij} g_{i}$), this tunability yields fast spin-flip transitions~\cite{doi:10.1126/science.1080880,PhysRevLett.120.137702,doi:10.1063/1.4858959,PhysRevApplied.16.054034,VenitucciElectricalmanipulationsemiconductor2018} and transversal interactions, while when $\textbf{B}\parallel \textbf{e}_i$ the interactions are longitudinal.  In the architectures examined, $\textbf{e}_i$ coincide with the main confinement axis~\cite{PhysRevResearch.2.033036} and a resonator field $E_r\parallel E$ yields 
\begin{equation}
\label{eq:short-dot-coupling}
\pmb{\gamma}_S=\frac{\mu_B}{2\hbar}\frac{\partial \underline{g}\cdot \textbf{B}}{\partial E }E_r  \approx  \frac{\alpha}{2el g_{i}}\frac{\partial g_{i}}{\partial E } \sqrt{\hbar Z_r }\omega_r \omega_B \textbf{e}_3\ .
\end{equation} 

In Fig.~\ref{fig:static-coupling}(b), we compare $\pmb{\gamma}_S$ in Si finFETs and in planar Ge for different directions of $\textbf{B}$. We assume that $\omega_B$  is constant and variations of $g_i$ at different $E$ are compensated by $\textbf{B}$. Importantly, this interaction is turned off at the sweet-spots with $\partial g_i/\partial E=0$,  where charge noise is suppressed to first order (marked with a square).
In contrast to $\pmb{\gamma}_L$, here  $\pmb{\gamma}_S \propto \omega_B\omega_r$, resulting in a lower enhancement of the longitudinal interactions by operating at $\omega_B\ll \omega_r$. Working in this regime however still  suppresses residual dispersive interactions arising, e.g. from small misalignments of $\textbf{B}$ from  $\textbf{e}_i$.
We emphasize that while typically smaller than $\pmb{\gamma}_L$, $\pmb{\gamma}_S$ conveniently results in significant interactions also in short quantum dots.

\paragraph*{Modulating the interactions.}

The spin-photon interactions are fully tunable. By changing electric potentials, the interactions can be turned on and off on-demand and can also be harmonically modulated, enabling efficient qubit readout~\cite{PhysRevLett.115.203601} and two-qubit gates~\cite{Royer2017fasthighfidelity,PhysRevB.96.115407}.

To tune $\pmb{\gamma}_L$,  we consider the protocol sketched in Fig.~\ref{fig:static-coupling}(c), where a gate (C) controls the length $l$ and the position of the  dot. 
In the off-state, the dot is short and C screens the electric field of the gate (R) connected to the resonator, suppressing $\pmb{\gamma}_L$. 
By squeezing the dot, the interactions are turned on and $\pmb{\gamma}_L$ is enhanced by orders of magnitude because $\pmb{\gamma}_L \propto l^4$ [Eq.~\eqref{eq:long-dot-coupling}] and $E_r$ is not screened by C.
Adiabatically  switching on $\pmb{\gamma}_L$ in a time $\tau_{on}\lesssim \omega_B/2\pi$ reduces errors and leakage during this protocol.
We emphasize that, as shown in the inset of Fig.~\ref{fig:static-coupling}(c),  the $g$-factor is also modified~\cite{adelsberger2021hole,PhysRevB.104.115425}  during the protocol, resulting in an additional phase accumulation in the qubit, that can be compensated for by single qubit gates or by appropriately modifying $\textbf{B}$.

In current experiments, the size of the dot is routinely varied in $\tau_{on}\sim 1$~ns~\cite{Fujita2017,Mills2019,Yoneda2021}, suggesting that this protocol could modulate $\pmb{\gamma}_L$  in ac. However, the strong non-linear dependence of $\pmb{\gamma}_L$ on $l$ might significantly distort the ac signal.
To reduce distortion, we propose instead to parametrically modulate the interactions via the $E$ dependence of $\pmb{\gamma}$. By superimposing a small ac field $\delta E(t)$ to the dc field $E$, we obtain
\begin{equation}
\pmb{\gamma}_{L,S}(t)\approx \pmb{\gamma}_{L,S}^0+ \pmb{\gamma}_{L,S}^1 \delta E(t)  \ .
\end{equation}

The dependence of the modulated interactions $\pmb{\gamma}_{L,S}^1$ on the dc field $E$ in Si finFETs is shown in Fig.~\ref{fig:static-coupling}(d).
Particularly relevant working points are the sweet spots where $\pmb{\gamma}_0=0$ (marked with squares and triangles). At those points, longitudinal interactions are  off at $\delta E(t)=0$, and the qubit lifetime is strongly enhanced because of a low susceptibility to charge noise.
The ac modulation of the interactions is much smaller than the static coupling, yielding $\pmb{\gamma}_{L}^1\sim 10\pmb{\gamma}_{S}^1\approx 0.5$~MHz, comparable to parametric longitudinal coupling in ST qubits~\cite{bottcher2021parametric}. We consider here large ac amplitudes $\delta E=0.1$~V/$\mu$m, and in analogy to before $\omega_r/2\pi=\omega_B/2\pi=5$~GHz, $V_r=20$~$\mu$V and $\alpha=0.4e$.
If the coupling is longitudinal, $\pmb{\gamma}_{L}^1$  is significantly enhanced by enlarging $\omega_r$, and   $\pmb{\gamma}_{L}^1\approx  10$~MHz is obtained for $\omega_r/2\pi\approx 25$~GHz.

\paragraph*{Signature of longitudinal interactions.}

\begin{figure}
\centering
\includegraphics[width=0.5\textwidth]{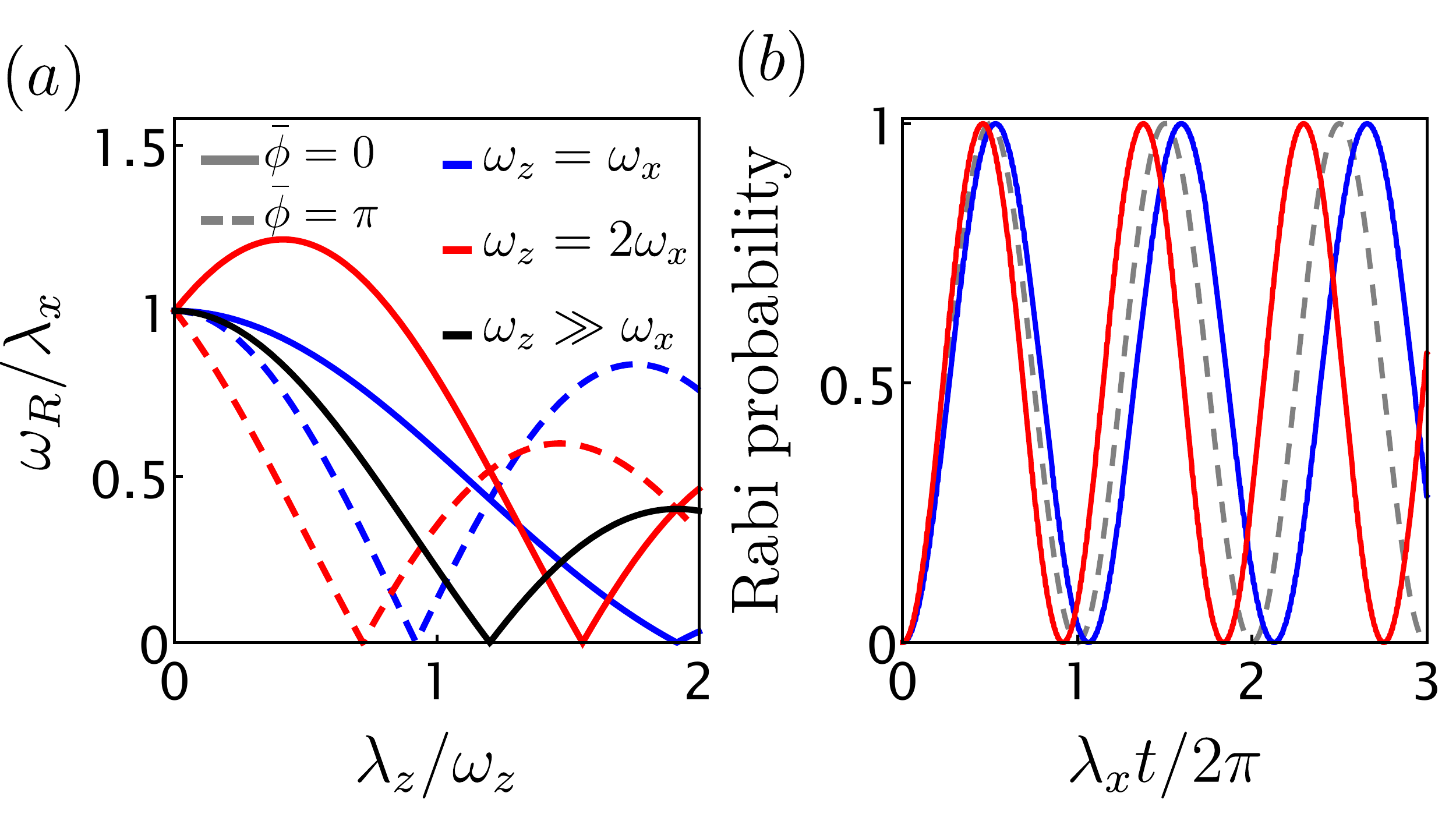}
\caption{\label{fig:measuring} 
Signatures of longitudinal interactions in the Rabi frequency $\omega_R$. In (a) we show $\omega_R$ against  $\lambda_z$. Here,  $\omega_B=\omega_x$, and we vary $\omega_z$ and the phase $\bar{\phi}=2 \omega_x(\phi+\pi)/\omega_z$. In~(b), we show the spin-flip probability in time, computed numerically from the Hamiltonian in Eq. ~\eqref{eq:drive-long}. We use $\omega_x=5\lambda_z=50\lambda_x$, $\bar{\phi}=0$,   and $\omega_z/\omega_x=1 (2)$ for the blue (red) line. For reference, we show with a dashed line the oscillations at $\lambda_z=0$.
}
\end{figure}

In striking contrast to transversal spin-photon  interactions, longitudinal interactions do not induce spin flip transitions nor qubit-dependent shifts of $\omega_r$, and thus characterizing them is a challenging task.
When both longitudinal and transversal interactions are present,  ac-modulated longitudinal interactions yield a measurable asymmetry in the qubit energy as a function of the detuning~\cite{bottcher2021parametric}, but this approach fails to measure static interactions. We propose  an alternative protocol based on the back-action of longitudinal interactions on externally driven spin rotations.
This protocol works in any system described by Eq.~\eqref{eq:Ham} and is not unique of hole spin qubits.

We consider an external driving of the qubit with frequency $\omega_x$ and amplitude $\lambda_x$, inducing Rabi oscillations. By  preparing the resonator in the coherent state $|\sqrt{\bar{n}} e^{i\omega_z t}\rangle$ with $\bar{n}$ photons, Eq.~\eqref{eq:Ham} at $\pmb{\gamma}\parallel \textbf{e}_3$ yields
\begin{equation}
\label{eq:drive-long}
H_M=\frac{\hbar\omega_B}{2}\sigma_3 + \hbar\lambda_x\cos(\omega_x t)\sigma_1 +\hbar\lambda_z\cos(\omega_z t+\phi)\sigma_3 \ ,
\end{equation}
with  $\lambda_z=2\sqrt{\bar{n}} |\pmb{\gamma}| $ and we include a phase difference $\phi$. 

At resonance $\omega_B=\omega_x$, and at $\lambda_z=0$, the state of the qubit rotates with Rabi frequency  $\omega_R=\lambda_x\sim 50-500$~MHz~\cite{Hendrickxsingleholespinqubit2019, Froning2021, Wang2022,hendrickx2020four,hendrickx2020fast,camenzind2021spin}. 
Finite values of $\lambda_z$ significantly alter the speed of spin precession. By moving to the  rotating frame with the operator $e^{-i\sigma_z  [\omega_x t/2+ \lambda_z\sin(\omega_z t+\phi)/\omega_z]}$, that accounts \textit{exactly}  for $\lambda_z$, and in the usual rotating wave approximation (RWA)   we find~\cite{SM}
\begin{equation}
\label{eq:long-drive-1}
\omega_{R}=\lambda_x  J_0\left(\frac{2\lambda_z}{\omega_z}\right)= \lambda_x \left[1-\frac{\lambda_z^2}{\omega_z^2}+\mathcal{O}\left(\frac{\lambda_z^4}{\omega_z^4}\right)\right] \ ,
\end{equation}
with $J_0$ being the Bessel function. The correction is quadratic in $\lambda_z/\omega_z$, but is still detectable in state-of-the-art architectures. 
For  realistic parameters $|\pmb{\gamma}|/2\pi= 100$~MHz, $\omega_z/2\pi= 5$~GHz, and in cavities with $\bar{n}=100$ photons, $\omega_R$ shifts by $\sim 15\%$ from $\lambda_x$.

Eq.~\eqref{eq:long-drive-1} is valid for arbitrary values of $\lambda_z$ and $\omega_z$, however, strikingly, at certain resonant frequencies $\omega_x=q \omega_z$, with $q=1/2$ or $q\in \mathbb{N}$ , the longitudinal corrections are modified by an extra phase-dependent term  and read
\begin{equation}
\label{eq:long-drive-2}
\omega_{R}^q=\lambda_x  \left| J_0\left(\frac{2\lambda_z}{\omega_z}\right)+e^{2iq (\phi+\pi)} J_{2q}\left(\frac{2\lambda_z}{\omega_z}\right)\right| \ .
\end{equation}
A detailed analysis of these resonances is provided in Sec.~II of~\cite{SM}.
A comparison between these cases for different values of $\phi$ is shown in Fig.~\ref{fig:measuring}, where we also confirm the shift of $\omega_R$ by numerically solving the Schr\"odinger equation with the Hamiltonian in Eq.~\eqref{eq:drive-long}.
In the weak driving limit $\lambda_z\lesssim \omega_z$, the sensitivity of $\omega_R$ on $\lambda_z$ is strongly enhanced at $q=1/2$ ($\omega_z=2\omega_x$), where $\omega_R^{q=1/2}/ \lambda_x\approx 1-\cos(\phi){\lambda_z}/{\omega_z}$, with a \textit{linear} dependence on $\lambda_z/\omega_z$. In long quantum dots, doubling $\omega_z\approx \omega_r$ enhances $\lambda_z/\omega_z\propto \omega_z$, see Eq.~\eqref{eq:long-dot-coupling}, yielding a maximal change of $\omega_R$ of $\sim 90\%$ for the same parameters used above. 
Interestingly, we also observe that at $\omega_z=2\omega_x$ and $\phi=\pi$, $\lambda_z$ enhances $\omega_R$ up to $22\%$.

\paragraph*{High-fidelity two-qubit gates.}

\begin{figure}
\centering
\includegraphics[width=0.5\textwidth]{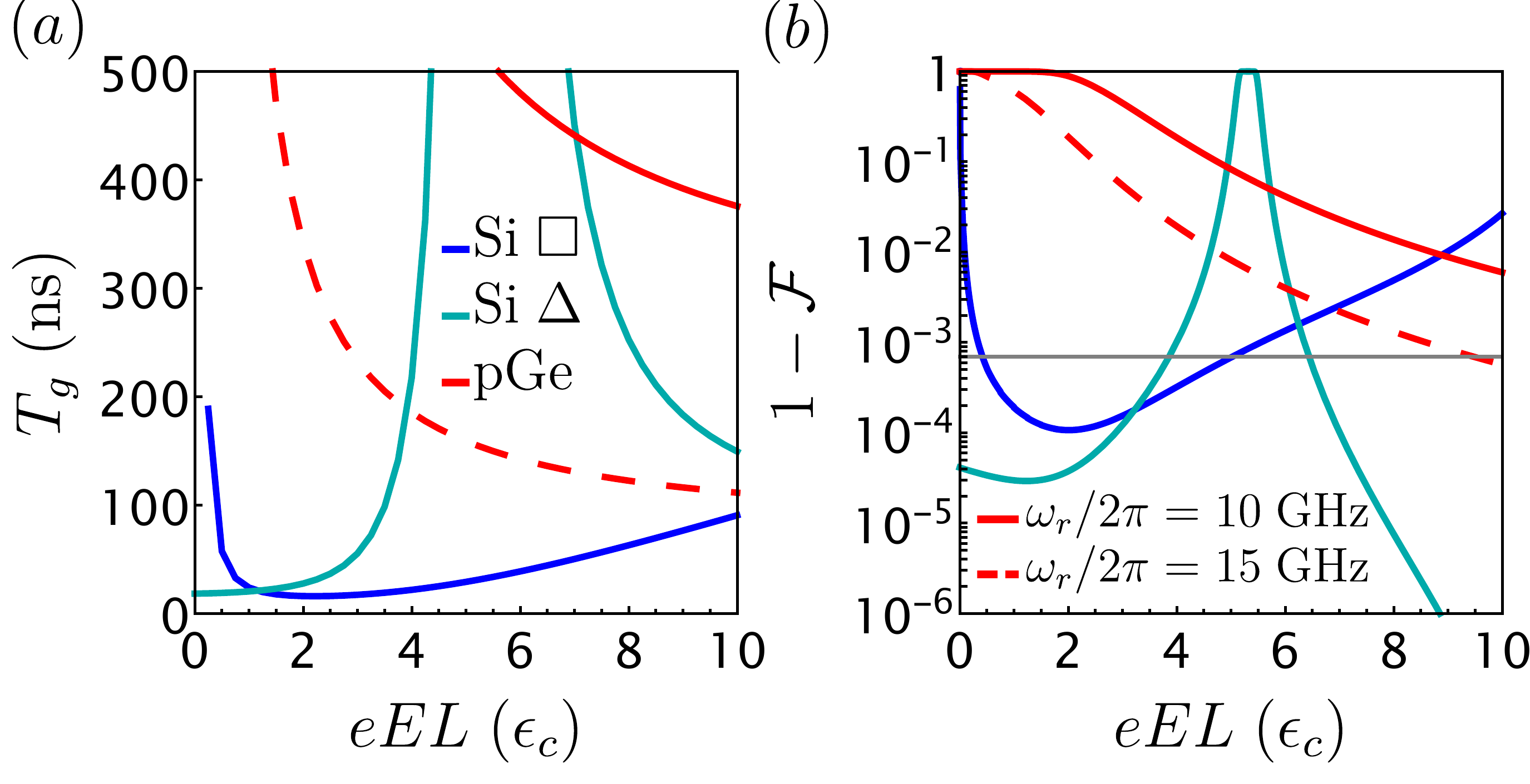}
\caption{\label{fig:fidelity} 
Controlled-Z gate of distant qubits.
We show   the gate time $T_g$ in (a) and the infidelity $1-\mathcal{F}$ (in logarithmic scale) in (b)  against $E$.
We analyze the long quantum dots described in Fig.~\ref{fig:static-coupling} (using the same color code), and resonators with $\alpha=0.3 e$, $Z_r=4$~k$\Omega$, and we mark with solid (dashed) lines the case $\omega_r/2\pi= 10 (15)$~GHz. The gray line indicates the surface code threshold $1-\mathcal{F}=7\times 10^{-4}$~\cite{PhysRevLett.109.180502}. We  consider a fixed qubit frequency $\omega_B/2\pi=5$~GHz and $1/f$ charge noise with $\alpha\bar{V}=5$~$\mu$eV. 
}
\end{figure}

When two qubits are longitudinally coupled with magnitude $\gamma_{1,2}$ to the same resonator with frequency $\omega_r$, the resonator mediates  effective Ising interactions $J \sigma_z^1\sigma_z^2$, with exchange $J=\gamma_1\gamma_2/\omega_r$~\cite{PhysRevB.91.094517}. 
A controlled-Z gate between these qubits $U_\text{cZ}=\text{diag}(1,1,1,-1)$ is  implemented by switching on $J$ for a time $T_g=\pi/4 J$ (up to single qubit rotations)~\cite{PhysRevA.57.120,SM}. The interactions $\gamma_{1,2}$ are turned on and off with the protocol sketched in Fig.~\ref{fig:static-coupling}(c). 

We consider two identical hole spin qubits encoded in the realistic long quantum dots discussed above, and in Fig.~\ref{fig:fidelity}(a) we show $T_g$. Strikingly, the large value of $\gamma$ results in fast gates with $T_g\sim 10-100$~ns, comparable to state-of-the-art superconducting qubits~\cite{PhysRevLett.125.240503,PhysRevLett.125.240502}.
Also, $T_g$ is significantly shortened at large $\omega_r$ because $J\propto \omega_r^3$. 

In contrast to alternative proposals~\cite{Royer2017fasthighfidelity,PhysRevB.103.035301,PhysRevB.96.115407}, our approach does not require an ac modulation of $\pmb{\gamma}$.
Here, we consider a square pulse of time $T_g$, and we neglect the turning on time $\tau_{on}$, as well as non-RWA corrections.
These effects are discussed in detail in Sec.~III of~\cite{SM}, where we present an exact solution also valid for smooth pulses.
A small value of $\tau_{on}\ll T_g$ only slightly affects $T_g$, and the non-RWA corrections are negligible if $\omega_r T_g\gg 1$. Strikingly, at $\omega_r T_g= 2\pi n$ with $n\in \mathbb{N}$, not only the non-RWA corrections vanish~\cite{Royer2017fasthighfidelity}, but our protocol works  for \textit{arbitrary} states of the resonator, including hot thermal states, suggesting that fast two-qubit gates could be executed at high temperatures. Demonstrating entangling gates at $4$~K,  compatible to current single qubit gates~\cite{camenzind2021spin} and cryo-CMOS~\cite{VandersypenInterfacingspinqubits2017,Gonzalez-Zalba2021,Xue2021,8993570,boter2021spider}, will pave the way towards large-scale quantum computers.

Gates mediated by longitudinal interactions have also high fidelity.
To examine the  fidelity  $\mathcal{F}$, we consider that  errors can arise from the decay rate $\kappa=\omega_r/Q$ of photons in resonators with quality factor $Q$. 
This noise channel yields a resonator-dependent dephasing of the qubits with time $T_r\approx\omega_r^2/\gamma^2\kappa\approx T_g Q$~\cite{Royer2017fasthighfidelity,PhysRevB.96.115407}.
However, in state-of-the-art resonators with $Q\sim 10^{4-5}$~\cite{PhysRevApplied.5.044004}, we estimate $T_r\gtrsim 0.1$~ms, much larger than the dephasing time of current hole spin qubits $T_2^\varphi\sim 0.1-10$~$\mu$s.
Moreover, residual transversal interactions caused by small misalignments of $\textbf{B}$ from $\textbf{e}_{so}$ with angle $\eta$ yield infidelities $1-\mathcal{F}\approx 0.8 (1+\bar{n})\eta^2\sim 10^{-3}-10^{-4}$ (see Sec.~IIID of~\cite{SM}) at $\eta\approx 1\%$ and for resonators with up to $\bar{n}= 10$ photons. Because this infidelity is suppressed by accurately aligning $\textbf{B}$, we focus now on the intrinsic qubit noise, resulting in~\cite{SM}
\begin{equation}
\mathcal{F}=e^{-(T_g/T_2^\varphi)^2} , \ \text{with} \ \frac{1}{T_2^\varphi}=\frac{\omega_B}{2 g \sqrt{\pi}}\bar{V}\frac{\partial g}{\partial V }  \sqrt{\ln\!\left(\frac{1}{\omega_{co}T_g}\right)} \ ,
\end{equation}
where $\omega_{co}\approx 1$~Hz is a frequency cut-off.
The gate infidelity $1-\mathcal{F}$ in different architectures is shown in Fig.~\ref{fig:fidelity}(b).
We consider dephasing caused by a $1/f$ charge noise~\cite{MAKHLIN2004315,PhysRevB.67.094510,PhysRevB.77.174509} with spectral function $S(\omega)=\bar{V}^2/|\omega|$ arising from random fluctuations of the gate potential. Assuming  that the control gate causes the largest fluctuations, we estimate ${\partial g}/{\partial V }\approx \alpha  \partial g/e\partial E l$, where typical values of $ \alpha \bar{V} \sim 0.1-10$~$\mu$eV~\cite{Yonedaquantumdotspinqubit2018,burkard2021semiconductor}, yields realistic $T_2^\varphi$ in the $\mu$s range.
Strikingly, in state-of-the-art devices we estimate high fidelities with $1-\mathcal{F}\sim 10^{-3}-10^{-4}$, above the surface code threshold~\cite{PhysRevLett.109.180502}, and pushing long-distance coupling to new speed and coherence standards. 

\paragraph*{Conclusion.}

We analyze  longitudinal spin-photon interactions in Si and Ge hole quantum dots. We show that these interactions are large, fully tunable and they can be harmonically modulated. These interactions are enhanced by  large qubit-resonator detuning, in striking contrast to dispersive interactions.
We propose protocols to quantify these interactions and to perform fast and high-fidelity two-qubit gates, that also work at high temperature. Engineering large longitudinal interactions in hole spin qubits will provide a significant step towards the implementation of a large-scale semiconductor quantum computer.\\

We thank L. Camenzind, S. Geyer, T. Patlatiuk and A. Kuhlmann for useful discussions.  
This work was supported as a part of NCCR SPIN funded by the Swiss National Science Foundation (grant number 51NF40-180604).

%

\clearpage
\newpage
\mbox{~}

\onecolumngrid

\begin{center}
  \textbf{\large Supplemental Material of  \\
Fully tunable longitudinal spin-photon interactions in Si and Ge quantum dots}\\[.2cm]
  Stefano Bosco, Pasquale Scarlino, Jelena Klinovaja and Daniel Loss\\[.1cm]
  {\itshape Department of Physics, University of Basel, Klingelbergstrasse 82, 4056 Basel, Switzerland}
   {\itshape Institute of Physics, Ecole Polytechnique F\'ed\'erale de Lausanne, CH-1015 Lausanne, Switzerland}\\
\end{center}

\setcounter{equation}{0}
\setcounter{figure}{0}
\setcounter{table}{0}
\setcounter{section}{0}

\renewcommand{\theequation}{S\arabic{equation}}
\renewcommand{\thefigure}{S\arabic{figure}}
\renewcommand{\thesection}{S\arabic{section}}
\renewcommand{\bibnumfmt}[1]{[S#1]}
\renewcommand{\citenumfont}[1]{S#1}

\title{Supplemental Material of \\
Fully tunable longitudinal spin-photon interactions in Si and Ge quantum dots}

\author{Stefano Bosco}
\email{stefano.bosco@unibas.ch}
\affiliation{Department of Physics, University of Basel, Klingelbergstrasse 82, 4056 Basel, Switzerland}
\author{Pasquale Scarlino}
\affiliation{Institute of Physics, Ecole Polytechnique F\'ed\'erale de Lausanne, CH-1015 Lausanne, Switzerland}
\author{Jelena Klinovaja}
\affiliation{Department of Physics, University of Basel, Klingelbergstrasse 82, 4056 Basel, Switzerland}
\author{Daniel Loss}
\affiliation{Department of Physics, University of Basel, Klingelbergstrasse 82, 4056 Basel, Switzerland}

\maketitle
\section*{Abstract}

We present  the theoretical model of hole spin qubits coupled to microwave resonators, and we derive the analytical expressions of the spin-photon interaction strength presented in the main text.
We also analyze in detail the modification of the Rabi frequency caused by a longitudinal driving, a mechanism that can be leveraged to measure longitudinal spin-photon interactions.  
Finally, we show general exact formulas of the time-evolution operator of a system comprising two qubits longitudinally coupled to the same resonator. Our results describe also longitudinal interactions with an arbitrary time dependence.  We examine gate time and gate fidelity when the interactions are smoothly turned on and when the resonator is in a thermal state. We show that fast and high-fidelity entangling gates can be implemented in state-of-the-art devices. 

\section{Hole spin qubits}
\label{sec:hole}
\subsection{Model of hole spin qubits}
Here, we describe in detail the microscopic model used to simulate spin qubits in hole quantum dots.
Hole nanostructures are well described by the Luttinger-Kohn Hamiltonian~\cite{WinklerSpinOrbitCoupling20031}
\begin{equation}
H_\text{LK}=\left(\gamma_1+\frac{5}{2}\gamma_2\right)\frac{\pmb{\pi}\cdot\pmb{\pi}}{2 m}-\frac{\gamma_2}{m}\pmb{\pi}^2\cdot\textbf{J}^2-2\frac{\gamma_3}{m}\left\{\pi_i,\pi_j\right\}\left\{J_i,J_j\right\}+ \text{c.p.} \ ,
\end{equation}
where c.p. means cyclic permutation, $\textbf{J}=(J_1,J_2,J_3)$ are spin 3/2 matrices, and $\pmb{\pi}=\textbf{p}+e\textbf{A}$ is the dynamical momentum with $\textbf{p}=-i\hbar\nabla$ being the canonical momentum and $\textbf{A}=(zB_y-yB_z/2,-zB_x+xB_z/2,0)$ is the vector potential. 
When simulating germanium (Ge) nanostructures, we use $\gamma_1=13.35$ and we work with the isotropic approximation, i.e. $\gamma_2=\gamma_3= 4.965$~\cite{WinklerSpinOrbitCoupling20031}; in this case, the results are independent of the growth direction of the wire and of the heterostructure. In silicon (Si) nanostructures, this approximation is not accurate and we consider fin field effect transistors (finFETs) with the fin aligned to $z\parallel[001]$ grown on a substrate aligned to $x\parallel[110]$. To model this case, we identify $p_z=p_3$ and  $p_{x,y}= (p_1\pm p_2)/\sqrt{2}$ ($J_z=J_3$ and $J_{x,y}=(J_1\pm J_2)/\sqrt{2}$). For Si, we use $\gamma_1=4.285$, $\gamma_2=0.339$ and $\gamma_3=1.446$~\cite{WinklerSpinOrbitCoupling20031}.
We define here the average hole mass $\bar{m}= m/\gamma_1$, where $m$ is the bare electron mass.

Different qubit designs are modelled by the confinement potentials and the strain.
In the main text, we study several  architectures.
Squeezed quantum dots in planar Ge heterostructures~\cite{PhysRevB.104.1154251} are modelled by considering a hard-wall confinement in the $y$-direction in a well with height $L$, and anisotropic parabolic confinement in $x$ and $z$ directions, with potential $V(x,z)=\hbar^2 (x^2/w^4+z^2/l^4)/2\bar{m}$, parametrized by the lengths $w<l$.
In these structures, we also include the strain caused by the mismatch of lattice constants in Ge and Si, that results in the additional energy $H_{S}=-\epsilon_s J_z^2$. We use $\epsilon_s=15$~meV, consistent with recent experiments in lightly strained heterostructures~\cite{lodari2021lightly1}.

Ge/Si core/shell nanowires~\cite{DRkloeffel11} are modelled by hard-wall confinement potential in a cross-section of circular shape, with radius $L$, and by harmonic potential $V(z)=\hbar^2 z^2/2\bar{m}l^4$ in the direction of the wire. In this case, we consider the strain energy $H_{S}=\epsilon_s J_z^2$, with $\epsilon_s=25$~meV, caused by the mismatch of lattice constants of Si and Ge~\cite{adelsberger2021hole1}.
In square~\cite{DRkloeffel31} and triangular~\cite{PRXQuantum.2.0103481} Si finFETs, we also consider hard-wall confinement with the appropriate cross-section with side $L$ and harmonic potential parametrized by $l$ in the $z$-direction.
In all these structures, we include an electric field in the $y$-direction, resulting in  $H_E=-eE y$.

We find the $g$-factors by applying a small magnetic field $\textbf{B}$, that modifies the momenta and induces a Zeeman energy $2\mu_B\kappa \textbf{B}\cdot \textbf{J}$. Here, $\kappa=3.41$ ($\kappa=-0.42$) for Ge (Si) and we neglect small corrections $\propto J^3_i$. By computing the derivative of $g$ with respect to $E$, we find the values of the spin-photon interaction $\gamma_S$.
The spin-photon interactions $\gamma_L$ are found by including the term $H_d=\omega_r d p (a^\dagger+ a)$ in the Luttinger-Kohn Hamiltonian, where $d=eE_r \bar{m}l^4 / \hbar^2$ is the dipole shift caused by the resonator electric field $E_r (a^\dagger+a)$. This term arises naturally by moving to a frame co-moving with the quantum dot center of mass~\cite{PhysRevB.104.1154251}. 
In the case of long-quantum dots, one can derive analytical expressions for $\gamma_L$, as shown in detail in the next section.

\subsection{Photon coupling to spins in long quantum dots}

We discuss here the interactions of a long quantum dot and a microwave resonator. Hole spin qubits in long quantum dots are well-approximated by a microscopic Hamiltonian, quadratic in the momentum operator $p$ in the long direction chosen to be along $z$-axis~\cite{PhysRevLett.127.1905011}. By including the spin-orbit interaction (SOI) $v$ and a resonator with frequency $\omega_r$, that shifts the position of the dot by an electric field $E_r$, we obtain
\begin{equation}
\label{eq:qubit-H}
H=\frac{p^2}{2m^*}+\frac{m^*\omega_o^2}{2}z^2+v p \pmb{\sigma}\cdot \textbf{e}_{so}+\hbar\omega_r a^\dagger a- eE_r z (a^\dagger+a) \ .
\end{equation}
Here, we introduce an effective wire mass $m^*$ (that typically varies from $\bar{m}$ by $10-20\%$) and an orbital frequency $\omega_o$. These quantities are related to the length $l$ by $m^*\omega_o^2=\hbar^2/\bar{m}l^4$.

While a magnetic field $\textbf{B}$ is required to split the qubit states, crucially,  the coupling of the dot to the resonator can be defined also at $\textbf{B}=0$. To observe this, we move to a frame co-moving with the dot by the translation $T=e^{-i p d (a^\dagger+a)/\hbar}$, where $d=eE_r/m^*\omega_o^2$ is characteristic shift of the wavefunction.
We then obtain
\begin{equation}
T^\dagger HT=\left(1+2\frac{\omega_r m^* d^2}{\hbar}\right)\frac{p^2}{2m^*}+\frac{m^*\omega_o^2}{2}z^2+vp\pmb{\sigma}\cdot \textbf{e}_{so}+\hbar\omega_r a^\dagger a-i\omega_r d p \left(a^\dagger-a\right)-\frac{m^*\omega_o^2d^2}{2}(a^\dagger+a)^2 \ . 
\end{equation}

The SOI is eliminated by the spin-dependent phase shift $S=e^{i \pmb{ \sigma}\cdot \textbf{e}_{so} z/l_{so}}$, with $l_{so}=\hbar/m^*v (1+2\omega_r m^*d^2/\hbar)$ being the spin-orbit length. We obtain
\begin{equation}
S^\dagger T^\dagger HTS=\left(1+2\frac{\omega_r m^* d^2}{\hbar}\right)\frac{p^2}{2m^*}+\frac{m^*\omega_o^2}{2}z^2+\hbar\omega_r a^\dagger a-i\omega_r d \left(p-\frac{\hbar}{l_{so}}\pmb{\sigma}\cdot \textbf{e}_{so}\right) \left(a^\dagger-a\right)-\frac{m^*\omega_o^2d^2}{2}(a^\dagger+a)^2 \ . 
\end{equation}
By projecting onto the lowest energy degenerate Kramers partners of the dot, one finds the spin-photon interaction Hamiltonian
\begin{equation}
\label{eq:Hinteractions}
H_{i}=i\left(a^\dagger-a\right)\hbar\omega_r \frac{d}{l_{so}}\pmb{\sigma}\cdot \textbf{e}_{so} \ ,
\end{equation}
 that is equivalent to the Hamiltonian provided in Eq.~(1) of the main text once the ladder operators are appropriately redefined. 
We emphasize that in our approach, no assumption is made on the size of the SOI, and thus our result is valid also for short $l_{so}\sim l$.
In the text, we neglect small corrections of the mass of the particle and of the frequency of the resonators. These terms do not introduce additional coupling terms between resonator and qubit, and thus do not alter qualitatively the discussion, and moreover, they are quantitatively small because they are $\propto \omega_r^2/\omega_o^2\sim 10^{-4}$.  The corrections to $H_i$ arising from different orbital eigenstates of the dot are also $\sim d^2/l^2$ and are negligible.

When a magnetic field is applied,  there is an additional contribution in Eq.~\eqref{eq:qubit-H} that takes the form
\begin{equation}
H_Z=\frac{\mu_B \textbf{B}\cdot\underline{g}_0(p)\cdot \pmb{\sigma}}{2} \ ,
\end{equation}
 with $\underline{g}_0(p)$ being a matrix of wire $g$-factors, that can depend on momentum. Because $[H_Z,T]=0$, the change of frame $T$ does not affect the Zeeman energy, and thus $H_Z$ does not produce additional coupling terms between the quantum dot and the microwave resonator. In contrast, the transformation $S$ shifts the momentum to $p\to p-\hbar/l_{so}\pmb{\sigma}\cdot\textbf{e}_{so}$ and  rotates the vector of Pauli matrices around the axis $\textbf{e}_{so}$~\cite{PhysRevLett.127.1905011}.
 For typical dependence in hole spin qubits $\underline{g}_0(p)=\delta_{ij}[\alpha_{i} -\beta_{i} p^2/\hbar^2+\mathcal{O}(p^4)]$ the transformation $S$ results in an additional length dependence of the quantum dot $g$-factor, and when projecting on the lowest energy Kramer partners of the  dot, one obtains
 $g=(\alpha-\beta/2l^2)e^{-l^2/l_{so}^2}$ if $\textbf{B}\perp \textbf{e}_{so}$ and $g=\alpha-\beta/2l^2$ if $\textbf{B}\parallel \textbf{e}_{so}$~\cite{PhysRevB.104.1154251}.
By defining $\omega_B= g \mu_B |\textbf{B}|/\hbar$ and by using Eq.~\eqref{eq:Hinteractions}, we obtain Eq.~(1) in the main text.

\section{Transversal and longitudinal driving}

Here, we derive the change of the Rabi frequency caused by a longitudinal drive. This mechanism enables to measure the longitudinal spin-photon coupling, as discussed in the main text.
We consider the Hamiltonian
\begin{equation}
H=\frac{\omega_B}{2}\sigma_z+\lambda_x \cos(\omega_x t+\phi_x) \sigma_x+\lambda_z \cos(\omega_z t+\phi_z) \sigma_z \ ,
\end{equation}
where to simplify the notation we set $\hbar=1$.
The polaron transformation
\begin{equation}
U_P=e^{-i\sigma_z\left[\frac{\omega_x t+\phi_x}{2}+z \sin\left(\omega_z t+\phi_z\right)\right] }  \ , \ \text{with} \ z=\frac{\lambda_z}{\omega_z} \ ,
\end{equation}
accounts exactly for  the longitudinal driving and modifies the rotating frame Hamiltonian as
\begin{equation}
\label{eq:H-driving}
\tilde{H}=U_P^\dagger H U_P-i U^\dagger_P \partial_t U_P=\frac{\Delta}{2}\sigma_z+\lambda_x \cos (\omega_x t +\phi_x) e^{i\left[\omega_x t+\phi_x+2z \sin\left(\omega_z t+\phi_z\right)\right] }  \sigma_+ +\text{h.c.}  \ .
\end{equation}

We introduce the detuning $\Delta=\omega_B-\omega_x$ between qubit and transverse driving field. At resonance $\Delta=0$, the Rabi frequency is, to a good approximation, given by  the time-independent component of the off-diagonal driving, that does not vanish in the rotating wave approximation (RWA). 
Without longitudinal driving, i.e. at $z=0$, the Rabi frequency is $\omega_R=\lambda_x$.
We emphasize that the resonance condition $\Delta=0$ imposes a constraint on the transversal driving frequency $\omega_x$ that needs to match the qubit frequency $\omega_B$, but there are no conditions on the longitudinal driving frequency $\omega_z$, that significantly influences the Rabi oscillations even when far detuned from $\omega_B$.

It is convenient to rewrite the off-diagonal driving term in Eq.~\eqref{eq:H-driving} by using the relation $e^{i x\sin(t)}=\sum_{n=-\infty}^\infty e^{i n t} J_{n}(x) $, with $J_n(x)$ being Bessel functions, resulting in
\begin{equation}
\label{eq:derivation_Fourier}
\lambda_x\cos (\omega_x t +\phi_x) e^{i\left[\omega_x t+\phi_x+2z \sin\left(\omega_z t+\phi_z\right)\right] } =\frac{\lambda_x}{2} \sum_{n=-\infty}^\infty  J_n(2z)e^{i n (\omega_z t+\phi_z)}\left( 1+ e^{2i(\omega_x t+\phi_x)} \right) \ .
\end{equation}

By discarding fast rotating terms in the RWA, one can generally neglect the second term  in the right hand side of Eq.~\eqref{eq:derivation_Fourier}, and  only the $n=0$ element of first term in the sum does not vanish. In this case, we obtain the Rabi frequency
\begin{equation}
\label{eq:Rabi-gen}
\omega_R= \lambda_x J_0(2 z) = \lambda_x \left(1- z^2\right) +\mathcal{O}\left(z^4\right) \ .
\end{equation}
The longitudinal drive renormalizes the Rabi frequency from its original value by a quadratic correction in $z$. Strikingly, because our theory is valid for arbitrary values of $z$ and $J_0(z)$ is an oscillating function vanishing at $z=1.2, \ 2.8,\cdots$, we  predict that Rabi oscillations can be suppressed by large longitudinal interactions.

Corrections to Eq.~\eqref{eq:Rabi-gen} arise when the driving frequencies $\omega_x$ and $\omega_z$ are appropriately chosen, such that the second term in the parenthesis in the right hand side of Eq.~\eqref{eq:derivation_Fourier} contributes in the RWA. 
In particular,  when $\omega_x/\omega_z= q$, with $q$ being a positive integer number or $q=1/2$, we find that the Rabi frequency acquires an additional correction dependent on the phase difference $\phi=\phi_x-q (\phi_z+\pi)$, and
\begin{equation}
\omega_R= \lambda_x \left|J_0(2 z)+e^{2i\phi}J_{2q}(2 z)\right|=\lambda_x\sqrt{J_0(2z)^2+J_{2q}(2z)^2+2J_0(2z)J_{2q}(2z)\cos(2\phi)} \ .
\end{equation}
Here, we use the relation $J_{m}(z)=(-1)^mJ_{-m}(z)$. 
Because $J_{m}(z)= z^m/2^mm! +\mathcal{O}(z^{m+1})$, in the small longitudinal driving limit $z\ll 1$, the only significant contributions of the additional phase-dependent term occur at $q=1$ or at $q=1/2$, i.e. when $\omega_z=\omega_x$ or $\omega_z=2\omega_x$, respectively.
In these cases, one finds the Rabi frequencies
\begin{subequations}
\begin{align}
\omega_R(q=1)&= \lambda_x \left|J_0(2 z)+e^{2i(\phi_x-\phi_z)}J_{2}(2 z)\right| = \lambda_x \left[1-z^2\left(1-\frac{1}{2}\cos (2\phi_x-2\phi_z)\right)\right]+\mathcal{O}\left(z^4\right) \ \ , \\
\omega_R(q=1/2)&= \lambda_x \left|J_0(2 z)-e^{i(2\phi_x-\phi_z)}J_{1}(2 z)\right| = \lambda_x \left[1-z \cos (2\phi_x-\phi_z)\right]+\mathcal{O}\left(z^2\right) \ \ .
\end{align}
\end{subequations}
We note that at $q=1/2$, the Rabi frequency becomes linearly dependent on $z$, thus enhancing the effect of $z$ on $\omega_R$ when $z\ll 1$. Interestingly, in this case $z$ can also enable faster Rabi oscillations, with  $\omega_R$ that are $22\%$ larger than $\lambda_x$ at $z=0.41$ and $2\phi_x-\phi_z=\pi$.

\section{Effective two-qubit interactions }
Here, we analyze the exact time evolution of a system comprising two qubits longitudinally coupled to the same resonator. In particular, we examine the time-dependent effective qubit-qubit interactions mediated by the resonator, and we show how to use them to implement fast and high-fidelity two-qubit gates. 

\subsection{Time-dependent Ising coupling}

\begin{figure}
\centering
\includegraphics[width=0.7\textwidth]{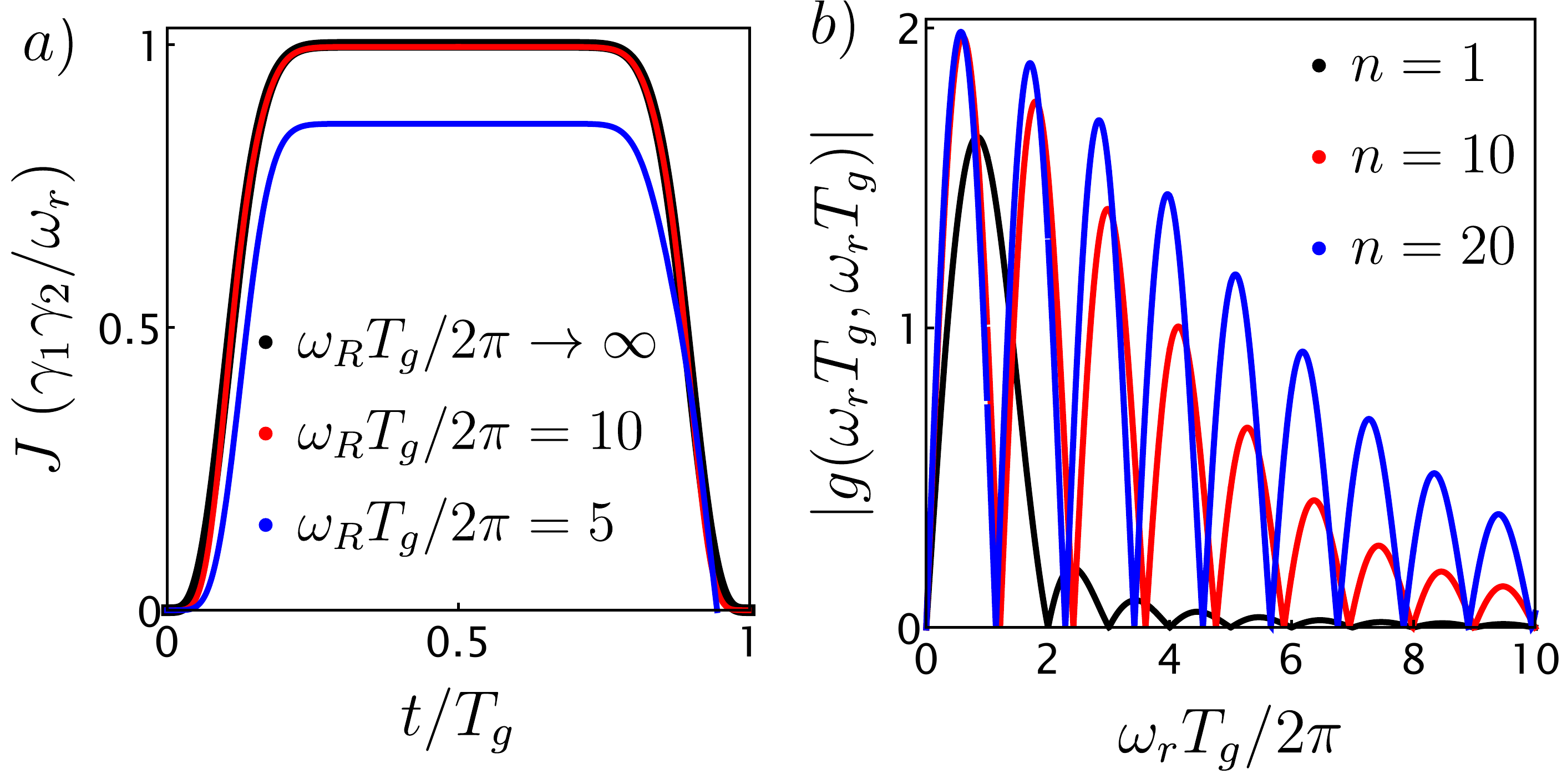}
\caption{\label{fig:pulse} Exact time evolution with smooth pulses. In a), we show the time-dependent Ising exchange interaction $J(t)$. We consider a pulse $\gamma_ {1,2}(t)=\gamma_{1,2}[1-\cos^{2n}(\pi t/T)]$ with $n=10$ and show that neglecting fast rotating terms (black line) is a well justified approximation at large values of $\omega_r T_g$. In b) we show the  absolute value of the displacement $\alpha(t)=-(\gamma_1\sigma_z^1+\gamma_2\sigma_z^2) g(\omega_r t,\omega_r T_g)/\omega_r$ after the gate time $T_g$ for different pulse abruptness $n$. We note that $\alpha(T_g)$ vanishes at certain resonator frequencies, that are well described by the condition $\omega_r T_g=2\pi n$ at $\omega_r T_g\gg 1$. At these frequencies the qubit and resonator are exactly decoupled after the two-qubit gate and the longitudinal coupling yield no back-action on the qubits. }
\end{figure}

Two qubits with frequencies $\omega_B^{1,2}$ longitudinally coupled to the same resonator with frequency $\omega_r$ are described by the Hamiltonian
\begin{equation}
\label{eq:full-H}
H=\frac{\omega_B^1\sigma_z^1+\omega_B^2\sigma_z^2}{2}+\omega_r a^\dagger a+ \Big[\gamma_1(t)\sigma_z^1+\gamma_2(t)\sigma_z^2\Big](a^\dagger+a) \ ,
\end{equation}
where to simplify the notation we set $\hbar=1$.
We analyze the relevant case where the couplings $\gamma_{1,2}$ are time-dependent. As discussed in the main text, in hole spin qubits, longitudinal interactions can be on-demand switched on and off  and can also be harmonically modulated by oscillating electric fields.

To remove the longitudinal interactions, we move to the polaron frame by the time-dependent displacement operator
\begin{equation}
\label{eq:unitary-pol}
D[\alpha(t)]=e^{\alpha(t)a^\dagger-\alpha(t)^*a}\ .
\end{equation} 
By using the relation $D[\alpha(t)]^\dagger\partial_tD[\alpha(t)]=\int_0^1 ds D^\dagger(s\alpha) [a^\dagger\partial_t{\alpha}-a\partial_t{\alpha}^*] D(s\alpha)$, and because longitudinal interactions commute with the qubit Hamiltonian, we find that when $\alpha$ satisfies
\begin{equation}
\label{eq:alpha_diff}
i\partial_t\alpha(t)=\omega_r\alpha(t)+\gamma_1(t)\sigma_z^1+\gamma_2(t)\sigma_z^2 \ ,
\end{equation}
the terms linear in the ladder operators vanish in the transformed Hamiltonian. 
Explicitly, we obtain the exact transformed Hamiltonian 
\begin{equation}
\label{eq:polaron-H}
\tilde{H}=D[\alpha(t)]^\dagger H D[\alpha(t)]-iD[\alpha(t)]^\dagger\partial_tD[\alpha(t)]=\frac{\omega_B^1\sigma_z^1+\omega_B^2\sigma_z^2}{2}+\omega_r a^\dagger a+  \frac{\omega_r }{2}|\alpha(t)|^2+\text{Re}[\alpha(t)]\Big[\gamma_1(t)\sigma_z^1+\gamma_2(t)\sigma_z^2\Big]\ .
\end{equation}
The last two terms result in an effective Ising interactions between the two qubits of the form $H_I=J(t)\sigma_z^1\sigma_z^2$.

When the couplings $\gamma_{1,2}$ are time independent, we find the time-independent displacement 
\begin{equation}
\alpha=-\frac{\gamma_1\sigma_z^1+\gamma_2\sigma_z^2}{\omega_r} \ , \ \text{and the Ising coupling} \ H_I=\frac{\gamma_1\gamma_2}{\omega_r}\sigma_z^1\sigma_z^2 \ ,
\end{equation}
in agreement with~\cite{PhysRevB.91.0945171,PhysRevB.96.1154071}.
Similar expressions can be derived when the interactions are slowly modulated and when the resonator frequency $\omega_r$ is much larger than the characteristic frequency of $\gamma_{1,2}(t)$.
This case is analyzed in the main text, where the interactions are turned on and off in a time $T_g\sim \omega_r/\gamma_1\gamma_2\gg 2\pi/\omega_r$. 
One can then approximate
\begin{equation}
\label{eq:rwa-J-alpha}
\alpha(t)\approx- \frac{\gamma_1(t)\sigma_z^1+\gamma_2(t)\sigma_z^2}{\omega_r} \ ,
\ \ \text{with \ time-dependent \ exchange \ interaction} \ \ \
J(t)\approx \frac{\gamma_1(t)\gamma_2(t)}{\omega_r} \ .
\end{equation}

To study in more detail the validity of this approximation, we consider the pulse $\gamma_{1,2}(t)=\gamma_{1,2} [1-\cos^{2n}(\pi t/T)]$~\cite{PhysRevB.96.1154071}, that approaches a square pulse at large values of the integer $n$, with a ramping  time $\tau_{on}\approx T_g/\sqrt{n}$. By solving the differential equation~\eqref{eq:alpha_diff} with initial condition $\alpha(0)=0$, one obtains $\alpha(t)=-(\gamma_1\sigma_z^1+\gamma_2\sigma_z^2)g(\omega_r t, \omega_r T_g)/\omega_r $, with
\begin{multline}
\label{eq:g-function}
g(t, T)=1-e^{-i t}-\frac{1+e^{\frac{2 i \pi  t}{T}}}{1-2 \pi  n/T} \cos^{2n}\left(\frac{\pi  t}{T}\right) \, _2F_1\left(1,1+n+\frac{T}{2 \pi };1-n+\frac{T}{2 \pi };-e^{\frac{2 i \pi  t}{T}}\right) \\ 
+\frac{2 e^{-i t}}{1-2 \pi  n/T} \, _2F_1\left(1,1+n+\frac{T}{2 \pi };1-n+\frac{T}{2 \pi };-1\right) \ ,
\end{multline}
and where $\, _2F_1$ is the hypergeometric function.
The resulting exchange coupling is then $J(t)=\gamma_1\gamma_2 \Big( 2[1-\cos^{2n}(\pi t/T_g)] \text{Re}[g(\omega_r t,\omega_r T_g)]-|g(\omega_r t, \omega_r T_g)|^2 \Big)/\omega_r$. At large values of $\omega_r T_g$, the exchange approaches $J(t)\approx\gamma_1\gamma_2 [1-\cos^{2n}(\pi t/T_g)]^2/\omega_r$, in agreement with Eq.~\eqref{eq:rwa-J-alpha}.
A comparison between the exact and approximate result at large $\omega_r$ is shown in Fig.~\ref{fig:pulse}a).

The results shown here can also be straightforwardly generalized to arbitrary pulse shapes and frequencies. 
In general, by considering a time-dependent coupling that is turned on at time $t=t_0$, we find that at a time $t$, the displacement $\alpha(t)$ is given by
\begin{equation}
\alpha(t)=-i e^{-i\omega_r t}\int_{t_0}^t ds e^{i\omega_r s}\Big[\gamma_1(s)\sigma_z^1+\gamma_2(s)\sigma_z^2\Big] \ ,
\end{equation}
and by defining the function $f(x,y)= [\gamma_1(x)\gamma_2(y) +\gamma_1(y)\gamma_2(x)]/2 $, the exchange interactions are
\begin{equation}
J(t)=\frac{1}{\omega_r}\left[\int_{\omega_r t_0}^{\omega_r t} ds \int_{\omega_r t_0}^{\omega_r t}ds' e^{i (s-s')} f(s/\omega_r,s'/\omega_r)-2\int_{\omega_r t_0}^{\omega_r t}ds\sin(\omega_r t-s)f( t,s/\omega_r)\right] \ .
\end{equation}
For example, by choosing a harmonic drive $\gamma_{1,2}(t)=\gamma_{1,2}\cos(\omega_d t)$  turned on at $t=0$, one obtains
\begin{equation}
\alpha(t)=-\frac{\gamma_1\sigma_z^1+\gamma_2\sigma_z^2}{2} \left(\frac{e^{-i t \omega _d}-e^{-i t \omega _r}}{\omega _r-\omega _d}+\frac{e^{i t \omega _d}-e^{-i t \omega _r}}{\omega _d+\omega _r}\right) \ .
\end{equation}
in agreement with~\cite{Royer2017fasthighfidelity1}.

\subsection{Two-qubit gates}

We now show that the exact time evolution of the system described by the Hamiltonian  $H$ in Eq.~\eqref{eq:full-H} yields a controlled phase gate between the two qubits, and that the fidelity of this gate is independent of the state of the resonator, even at finite temperature.

By using the displacement operator $D[\alpha(t)]$ defined in Eq.~\eqref{eq:unitary-pol}, we find that time evolution induced by $H$ is determined  by the unitary operator
\begin{equation}
\label{eq:U-tot}
U(t)=D[\alpha(t)]e^{-i \int_{t_0}^t \tilde{H}(\tau) d\tau}D[\alpha(t_0)]^\dagger =e^{-i [\sigma_z^1 \omega_B^1+\sigma_z^2 \omega_B^2](t-t_0)/2} e^{-i \sigma_z^1 \sigma_z^2 \int_{t_0}^t J(\tau) d\tau}D[\alpha(t)]  e^{-i \omega_r (t-t_0) a^\dagger a}D[\alpha(t_0)]^\dagger\ ,
\end{equation}
where $\tilde{H}$ is the Hamiltonian in Eq.~\eqref{eq:polaron-H}.
The first and second exponentials represent a single and two qubit rotation, respectively. Because  $\alpha(t)$ depends on $\sigma_z^{1,2}$, the last term describes the coupled dynamics of qubits and resonator.

Let us first neglect the last term in the time evolution operator, and by calling $\bar{J}(t)=\int_{t_0}^t J(\tau) d\tau$, we obtain 
\begin{equation}
\label{eq:U0-quibt}
U(t)\approx U_0(t)= e^{-i [\sigma_z^1 \omega_B^1+\sigma_z^2 \omega_B^2](t-t_0)/2} e^{-i \sigma_z^1 \sigma_z^2 \bar{J}(t)} \ .
\end{equation}
This unitary transformation is an entangling controlled phase gate, up to single qubit rotations. In particular,  by turning on the interactions for  a time $T_g$, satisfying $\bar{J}(T_g)=\pi/4$, $U_0(T_g)$ yields a controlled Z gate $U_\text{cZ}=\text{diag}(1,1,1,-1)$, up to a global phase factor and single qubit $Z$ rotations $e^{-i \sigma_z \beta_i}$ with angles $\beta_i=-(\omega_B^i T_g/2+\pi/4)$~\cite{PhysRevA.57.1201}.
By considering  a simple square pulse, one obtains the gate time $T_g^\text{sq}=\pi\omega_r/4 \gamma_1\gamma_2$. In the main text, we estimate $T_g^\text{sq}\sim 10-100$~ns in realistic devices, enabling fast two-qubit gates.
If we  consider the trigonometric pulse shape $J(t)=\gamma_1\gamma_2[1-\cos^{2n}(\pi t/T)]^2/\omega_r$, see Fig.~\ref{fig:pulse}, the gate time  increases as
\begin{equation}
T_g^\text{cos}=T_g^\text{sq}\left(1+\frac{\Gamma \left(2 n+\frac{1}{2}\right)}{\sqrt{\pi }(2 n)!}-\frac{2 \Gamma \left(n+\frac{1}{2}\right)}{\sqrt{\pi }n!}\right)^{-1} \ .
\end{equation}
The correction to $T_g$ is largest at $n=1$, where $T_g^\text{cos}/T_g^\text{sq}=8/3$ and it approaches 1 as $n$ increases: for example at $n=10$, $T_g^\text{cos}/T_g^\text{sq}\approx 1.3$. In the main text, we neglect this small correction.\\

The last exponential in Eq.~\eqref{eq:U-tot} introduces additional spin-dependent dynamics in the resonator, resulting in small non-RWA corrections. These corrections could yield an unwanted residual entanglement between the qubit and the resonator at the end of the operation, however, by appropriately choosing the gate time $T_g$ and the resonator frequency $\omega_r$ these small corrections can be exactly cancelled.

To illustrate this, we assume at first that at $t=t_0$, the resonator is prepared in a coherent state $|\beta\rangle$ and the qubits are in a general state $|\Psi\rangle=\sum_{s_1s_2} c_{s_1s_2}|s_1s_2\rangle$, with normalized coefficients $c_{s_1s_2}$; also $\alpha(t_0)=0$. At time $t>t_0$, the resonator state remains coherent and evolves in the spin-dependent coherent state
\begin{equation}
D[\alpha(t)]  e^{-i \omega_r (t-t_0) a^\dagger a}D[\alpha(t_0)]^\dagger|\Psi\rangle|\beta\rangle=\sum_{s_1s_2} c_{s_1s_2}|s_1s_2\rangle|\beta e^{-i\omega_r (t-t_0)}+\alpha_{s_1s_2}(t)\rangle \ ,
\end{equation}
where we used the eigenvalue equation $\alpha(t)|s_1s_2\rangle=\alpha_{s_1s_2}(t)|s_1,s_2\rangle$.
To suppress unwanted couplings, the gate time $T_g$ and the resonator frequency $\omega_r$ needs to satisfy $\alpha(T_g+t_0)=0$, such that the final state factorizes and $D[\alpha(T_g+t_0)]  e^{-i \omega_r T_g a^\dagger a}D[\alpha(t_0)]^\dagger|\Psi\rangle|\beta\rangle=|\Psi\rangle|\beta e^{-i\omega_r T_g}\rangle$. This condition is always respected in the RWA because in this approximation $\alpha(t)\propto \gamma_1(t)\gamma_2(t)$, see Eq.~\eqref{eq:rwa-J-alpha}, and  $\gamma_{1,2}(T_g+t_0)=0$. Even when the RWA is not applicable, we note that $\alpha(T_g+t_0)$  vanishes  at certain resonator frequencies, as shown in  Fig.~\ref{fig:pulse}b). In particular, when $\omega_r T_g\gg 1$, the condition $\alpha(T_g+t_0)=0$ is satisfied  at $\omega_r T_g=2\pi n$. Working at these resonator frequencies thus provides a general way to exactly remove  residual resonator-qubit interactions at the end of the operations~\cite{Royer2017fasthighfidelity1}. 

This important result holds also when the initial state of the resonator is a mixed state.
For example, we now analyze explicitly what happens when at time $t=t_0$ the resonator is in a thermal state and the initial density matrix $\rho$ of the system reads
\begin{equation}
\rho(t=t_0)=\rho_1\otimes\rho_2\otimes \frac{e^{-\omega_r a^\dagger a/T }}{\mathcal{Z}} \ , \ \text{with partition function} \ \mathcal{Z}=\text{Tr}(e^{-\omega_r a^\dagger a/ T })=\frac{1}{1-e^{-\omega_r/ T }} \ .
\end{equation}
We introduce the temperature $T$, and to simplify the notation, we set the Boltzmann constant $k_B=1$.
The initial qubit density matrices are generally given by $\rho_{i}=\sum_{s_i,s_i'}c_{s_is_i'}|s_i\rangle \langle s_i' |$, with normalized coefficients $c_{s_is_i'}$, and $i=\{1,2\}$. By using $\sigma_z^i|s_i\rangle=s_i|s_i\rangle$, we find that the system evolves as
\begin{align}
\rho(t)=&U(t)\rho(t_0) U(t)^\dagger =\sum_{s_1s_1's_2s_2'}c_{s_1s_1'}c_{s_2s_2'}e^{i [(s_1'-s_1) \omega_B^1+(s_2'-s_2) \omega_B^2](t-t_0)/2} e^{i (s_1' s_2'-s_1 s_2) \bar{J}(t)}|s_1s_2\rangle \langle s_1's_2'| \\ 
& \times 
D[\alpha_{s_1s_2}(t)] e^{-i \omega_r (t-t_0) a^\dagger a}D[\alpha_{s_1s_2}(t_0)]^\dagger \frac{e^{-\omega_r a^\dagger a/T }}{\mathcal{Z}} D[\alpha_{s_1's_2'}(t_0)] e^{i \omega_r (t-t_0) a^\dagger a}D[\alpha_{s_1's_2'}(t)]^\dagger\ .
\end{align}
The first line of this equation describes the target controlled-phase operation (up to single qubit rotations), while the last line of this equation describes the time evolution of the state of the resonator caused by the longitudinal spin-photon interactions. Remarkably, in analogy to the pure state case,  if the gate time $T_g$ satisfies $\alpha(T_g+t_0)=0$ [at $\omega_r T_g= 2\pi n$, see Fig.~\ref{fig:pulse}b)],  the  resonator state  evolves  back into its initial state after the operation, yielding no back-action on the qubits. This property indicates that the state of the resonator does not need any preparation and that the proposed long-range entangling gate  works well also at high temperatures.

\subsection{Fidelity of two-qubit gate}

We now discuss the fidelity of two-qubit gates.
We first note that  photons in state-of-the-art high-impedance resonators have a typical lifetime $1/\kappa\approx Q/\omega_r$, where the resonator quality factor is $Q\approx 10^{4}-10^5$~\cite{PhysRevApplied.5.0440041}. Because $1/\kappa$ is much longer than the typical decoherence time  $T_2^\varphi\sim 0.1-10$~$\mu$s of current hole spin qubits~\cite{Hendrickxsingleholespinqubit20191,hendrickx2020four1,hendrickx2020fast1,camenzind2021spin1},  the fidelity of the two qubit gates is limited by the qubit noise. In particular, by following~\cite{Royer2017fasthighfidelity1}, we estimate that the typical resonator-induced dephasing  time  $T^\varphi_r\approx \omega_r^2/\kappa \gamma^2\approx T_g Q\approx 0.1$~ms ($Q=10^4$,  $T_g^{-1}=100$~MHz) is orders of magnitude larger than $T_2^\varphi$.

To model the decoherence of the qubit , we consider two random  variables $h_{1,2}(t)$ coupled to the two qubits by the noisy Hamiltonian~\cite{PhysRevB.67.0945101,PhysRevB.77.1745091}
\begin{equation}
H_N=H+\frac{h_1(t)\sigma_z^1+h_2(t)\sigma_z^2}{2} \ ,
\end{equation}
where $H$ is given in Eq.~\eqref{eq:full-H}. 
Because in current experiments the relaxation time $T_1$ is much longer than $T_2$, in our model we focus only on pure dephasing and neglect transversal noise terms $\propto \sigma_{x,y}$.
We analyze two quibts that are rather distant from each other, and thus we do not include correlated noise $\sim \sigma_z^1\sigma_z^2$ in the lab frame. Moreover,  the  displacement $D[\alpha(t)]$ in Eq.~\eqref{eq:unitary-pol} commutes with the noise, and thus in this model there is no correlated noise arising from the resonator. The exact transformed Hamiltonian in the polaron frame is
\begin{equation}
\tilde{H}_N=\tilde{H}+\frac{h_1(t)\sigma_z^1+h_2(t)\sigma_z^2}{2} \ ,
\end{equation}
with $\tilde{H}$ being given in Eq.~\eqref{eq:polaron-H}. We note that because  the photon and qubit degrees of freedom are decoupled in $\tilde{H}$, the noise of the qubit is independent of the state of the resonator, in striking contrast to transversal spin-photon interactions. 

The time evolution of the system in the lab frame is  given by 
\begin{equation}
U_N(t)=D[\alpha(t)]e^{-i \int_{t_0}^t \tilde{H}_N(\tau) d\tau}D[\alpha(t_0)]^\dagger =U(t) e^{-i \int_{t_0}^t [h_1(\tau)\sigma_z^1+h_2(\tau)] d\tau/2}  \ .
\end{equation}
Here $U(t)$ is the target unitary evolution of the system, given in Eq.~\eqref{eq:U-tot}, that enables two-qubit entangling gates (up to single qubit operations).
By following standard derivations~\cite{PhysRevB.67.0945101,PhysRevB.77.1745091} and considering a Gaussian noise, where the variables $x_{i}(t)=\int_{t_0}^td\tau h_i(\tau)/2$ are Gaussian distributed, we average the time-evolution operator over a Gaussian probability distribution with standard deviation $\sigma_i$ and zero mean value, yielding
\begin{equation}
\label{eq:noise-TEO}
\bar{U}_N(t)=U(t) \int dx_1 \frac{e^{-x^2_1/2\sigma^2_1}}{\sqrt{2\pi}\sigma_1}e^{-i x_1\sigma_z^1} \int dx_2 \frac{e^{-x^2_2/2\sigma^2_2}}{\sqrt{2\pi}\sigma_2}e^{-i x_2\sigma_z^2} =U(t) e^{-[\sigma^2_1(t)+\sigma_2^2(t)]/2} \ .
\end{equation}
The covariance of $x_{i}(t)$ is given by
\begin{equation}
\sigma^2_i(t)=\frac{1}{\pi}\int_{0}^{\infty} d\omega \frac{\sin^2(\omega t/2)}{\omega^2} S_i(\omega)  \ , 
\end{equation}
where $S_i(\omega)=\int_{-\infty}^{\infty}dt e^{i\omega t} \langle h_i(t)h_i(0)\rangle$ is the spectral function of the noise on qubit $i$,  and the prefactor in the integration is the usual filter function of free induction decay experiments~\cite{PhysRevB.67.0945101,PhysRevB.77.1745091}.
For example, assuming a typical $1/f$ charge noise with $S_{i}(\omega)= A_{i}^2/|\omega|$ parametrized by the energies $A_i$, and introducing a small frequency cut-off  $\omega_{co}$, one obtains the approximate Gaussian decay
\begin{equation}
\bar{U}_N(t)\approx U(t)e^{-t^2[(T_2^\varphi)^{-2}_1+(T_2^\varphi)^{-2}_2]/2} \ , \ \text{with} \  \frac{1}{(T_2^\varphi)_i}=\frac{A_i}{2\sqrt{\pi}\hbar} \sqrt{\ln\!\left(\frac{1}{\omega_{co}t}\right)}\approx \frac{A_i}{2\sqrt{\pi}\hbar} \ .
\end{equation}
We note that the total time evolution of the system decays as a Gaussian function with a rate that is set by the root mean square of the dephasing rates $1/(T_2^\varphi)_i$ of the individual qubits. \\ 

\begin{figure}
\centering
\includegraphics[width=0.3\textwidth]{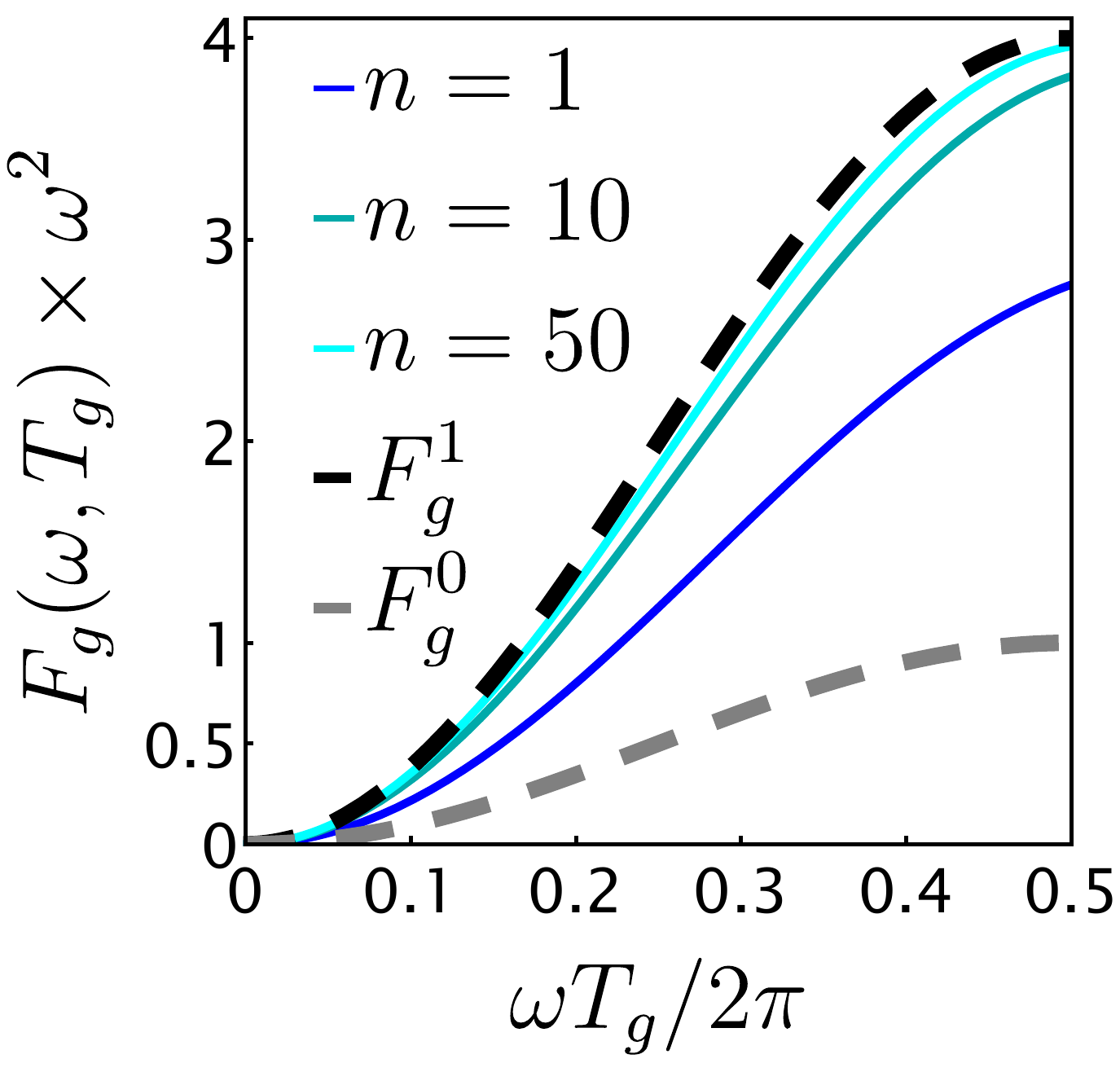}
\caption{\label{fig:noise} Filter function $F_g$ of the noise in the presence of a time-dependent noise sensitivity $\partial g(t)/\partial V$. We show the filter function of the noise $F_g$ in Eq.~\eqref{eq:filter-funct} evaluated at the gate time $T_g$ as a function of frequency $\omega$ and for different ramping times $\tau_{on}\approx T_g/\sqrt{n}$. For the plot we use $\delta g_0=1$ and $\delta g_1=2$. With dashed gray and black lines we show the limiting cases of constant noise sensitivity $F_g^0=\delta g_0^2 \sin(\omega t/2)^2/\omega^2$ and $F_g^1=\delta g_1^2 \sin(\omega t/2)^2/\omega^2$, respectively. The filter function at small frequencies is bounded between those limiting cases and it approaches the black dashed line ($F_g^1$) at sufficiently fast $\tau_{on}$. }
\end{figure}

More specifically, in the main text we propose a protocol to adiabatically turn on the longitudinal interactions by electrically tuning the length of the quantum dot  in a time $\tau_{on}$ that satisfies $2\pi/\omega_B\gg\tau_{on}\gg T_g $. 
Because of the large SOI, tuning the length of the dot results in a coherent modulation of the $g$-factor, and in an extra phase accumulation in the qubits, see Sec.~\ref{sec:hole}. This effect can be measured and corrected, however, the time-dependent modulation of the $g$-factor could also affect the noise and $T_2^\varphi$. 
In particular, the noise can be modelled  as 
\begin{equation}
h_1(t)=h_2(t)=\mu_B B \frac{\partial g(t)}{\partial V}  \delta V(t) \ , 
\end{equation} 
where $g(t)$ accounts for the coherent modulation of $g$ and $\delta V(t)$ is a random potential fluctuation, with spectral noise $S_V(\omega)$ and autocorrelation 
\begin{equation}
\langle \delta V(t) \delta V(0) \rangle=\int_{-\infty}^{\infty}\frac{d\omega}{2\pi}e^{-i\omega t} S_V(\omega) \ .
\end{equation}
The derivations presented above, and in particular Eq.~\eqref{eq:noise-TEO}, are still valid, but with covariance
\begin{equation}
\sigma^2(t)=\frac{(\mu_B B)^2}{\pi}\int_{0}^{\infty}d\omega F_g(\omega, t)  S_V(\omega)  \ , \ \text{and} \  F_g(\omega, t)=\left|\frac{1}{2} \int_{t_0}^t d\tau \frac{\partial g(\tau)}{\partial V} e^{i\omega \tau} \right|^2 \ ,
\end{equation}
that includes the time-dependent noise sensitivity  ${\partial g(t)}/{\partial V}$ and where $F_g(\omega, t)$ is the filter function.

To understand the effect of this time-dependent noise sensitivity, let us consider the insightful example
\begin{equation}
 \frac{\partial g(t)}{\partial V}=  \delta g_0+ (\delta g_1 -\delta g_0) [1-\cos^{2n}(\pi t/T_g)] \ ,
\end{equation}
where the noise sensitivity smoothly interpolates between the value $\delta g_0$ at $t=0$ and the value $\delta g_1$ in a time $T_g$. This function well describes the protocols we study in the main text and in the previous sections. In this case, the filter function is
\begin{equation}
\label{eq:filter-funct}
F_g(\omega, t)=\frac{\left|  (1-e^{-i\omega t})\delta g_0+  g(\omega t,\omega T_g)(\delta g_1-\delta g_0) \right|^2}{4\omega^2} \ ,
\end{equation}
where the function $g(t,T)$ is given in Eq.~\eqref{eq:g-function}.

In Fig.~\ref{fig:noise}, we show a comparison between filter functions  $F_g$ in the relevant small frequency case for different ramping times $\tau_{on}\approx T_g/\sqrt{n}$. We observe that the filter function is bounded between the limiting cases $F_g^0=\delta g_0^2 \sin^2(\omega t/2)/\omega^2$ and $F_g^1=\delta g_1^2 \sin^2(\omega t/2)/\omega^2$ obtained by considering constant noise sensitivity. As the ramping time decreases and the pulse is sharper (at large $n$), the relevant noise sensitivity is $\delta g_1$ and the variation from $F_g^1$ become negligible at small frequencies.
For this reason, in the main text, where we analyze a sharp pulse, we use the filter function $F_g^1=\delta g_1^2 \sin(\omega t/2)^2/\omega^2$, yielding
\begin{equation}
\frac{1}{T_2^\varphi}=\omega_B \frac{\delta g_1 \bar{V}}{2g\sqrt{\pi}} \sqrt{\ln\!\left(\frac{1}{\omega_{co}t}\right)} \ ,
\end{equation}
 for a $1/f$ noise with $S_V(\omega)=\bar{V}^2/|\omega|$.

By defining the average fidelity $\mathcal{F}$ of the entangling gate as the average variation of $\bar{U}_N(t)$ from the targeted time evolution operator $U(t)$ at time $t=T_g$, we obtain the formula 
\begin{equation}
\label{eq:fid-noise}
\mathcal{F}=\text{Tr}[U^\dagger(T_g)\bar{U}_N(T_g)]=e^{-(T_g/T_2^\varphi)^2} \ ,
\end{equation}
used in the main text. Here, the trace is normalized over the dimension of the Hilbert space. By considering the values of gate time  $T_g\approx 10-100$~ns estimated in the main text and a typical experimental value $T_2^\varphi \approx 1$~$\mu$s, we expect high-fidelity operations, with $1-\mathcal{F}\approx 10^{-2}-10^{-4}$.

\subsection{Residual transversal interactions}

We now discuss the effect of residual transversal interactions on the two-qubit gate.
These interactions can arise from the misalignment of the magnetic field from the desired direction, e.g. the spin-orbit direction or the principal axis of the $g$-tensor, and yield the additional energy
\begin{equation}
H_T= \eta \gamma(t) (\sigma_x^1+\sigma_x^2)(a^\dagger+a) \ ,
\end{equation}
where $\eta$ parametrizes the misalignment of the field.
For simplicity, here we consider two identical qubits with the same $\gamma(t)$, $\eta$, and $\omega_z$, however, our treatment can be straightforwardly generalized to different qubits.

By moving to the interaction picture with the unitary transformation $U(t)$ in Eq.~\eqref{eq:U-tot} and by setting $t_0=0$ without loss of generality, the interaction picture Hamiltonian is
\begin{align}
\tilde{H}_T&=U(t)^\dagger  H_T U(t)=\eta \gamma(t) U_0(t)^\dagger e^{i\omega_r t a^\dagger a}D[\alpha(t)]^\dagger (a^\dagger+a)(\sigma_x^1+\sigma_x^2)D[\alpha(t)]e^{-i\omega_r t a^\dagger a}U_0(t) \\
&= \eta \gamma(t) U_0(t)^\dagger e^{i\omega_r t a^\dagger a} \left[a^\dagger+a-\frac{2\gamma(t)}{\omega_r}(\sigma_z^1+\sigma_z^2)\right]\left[(\sigma_+^1+\sigma_+^2)D\!\left(\frac{2\gamma(t)}{\omega_r}\right)+(\sigma_-^1+\sigma_-^2)D\!\left(\frac{2\gamma(t)}{\omega_r}\right)^\dagger \right] e^{-i\omega_r t a^\dagger a}U_0(t) \ , \\
& \approx \eta \gamma(t) U_0(t)^\dagger e^{i\omega_r t a^\dagger a} \left[(\sigma_+^1+\sigma_+^2)a +(\sigma_-^1+\sigma_-^2) a^\dagger \right] e^{-i\omega_r t a^\dagger a}U_0(t) \ , 
\end{align}
with $U_0(t)$ being the operator acting only on the qubits defined in Eq.~\eqref{eq:U0-quibt}, and in the second line, we use $\alpha(t)= -\gamma(t)(\sigma_z^1+\sigma_z^2)/\omega_r$. The third line shows the leading correction at small values of $\gamma(t)/\omega_r$ obtained in the RWA.
The unitary evolution operator is then modified as
\begin{equation}
U_{T}(t)=U(t) \mathcal{T}\!\exp\left[-i\int_0^t d \tau \tilde{H}_T(\tau)\right]U(0)^\dagger 
= U(t)\left[1 - i \int_0^t d \tau \tilde{H}_T(\tau) - \int_0^t d \tau \int_0^\tau d \tau' \tilde{H}_T(\tau)   \tilde{H}_T(\tau')+\mathcal{O}(\eta^3)\right] \ .
\end{equation}

We quantify the effect of these interactions on the two qubit gate by studying how the final state differs from the target state. In analogy to Eq.~\eqref{eq:fid-noise}, we define the average fidelity
\begin{equation}
\label{eq:fid-transv}
\mathcal{F}_T=\frac{1}{d}\langle \beta| \text{Tr}_Q[U^\dagger(T_g)U_T(T_g)]|\beta\rangle=1 -\frac{1}{d} \int_0^{T_g} d \tau \int_0^\tau d \tau'\langle \beta| \text{Tr}_Q[\tilde{H}_T(\tau)\tilde{H}_T(\tau')] |\beta\rangle+\mathcal{O}(\eta^3) \ ,
\end{equation}
and we use $\text{Tr}_Q[\tilde{H}_T(\tau)]=0$. In striking contrast to Eq.~\eqref{eq:fid-noise}, $\mathcal{F}_T$ depends on the state of the resonator, and thus we trace over the qubit degrees of freedom (normalized by $d=4$) and we consider a resonator prepared at $t=0$ in the coherent state $|\beta\rangle$.
In the RWA, we find
  \begin{align}
\label{eq:fid-transv_RWA}
\mathcal{F}_T&\approx 1 -\eta^2 \int_0^{T_g} d \tau \gamma(\tau) \int_0^\tau d \tau' \gamma(\tau') \cos[2 (J(\tau)-J(\tau'))]\langle \beta|\left[ e^{i\omega_B(\tau-\tau')}a(\tau)a^\dagger(\tau') +  e^{-i\omega_B(\tau-\tau')}a^\dagger(\tau)a(\tau')\right]  |\beta\rangle \\
&=1 -\eta^2 \int_0^{T_g} d \tau \gamma(\tau) \int_0^\tau d \tau' \gamma(\tau') \cos[2 (J(\tau)-J(\tau'))]\left[ e^{-i(\omega_r-\omega_B)(\tau-\tau')} +|\beta|^2  e^{i(\omega_r-\omega_B)(\tau-\tau')}\right] \ ,
\end{align}
where $a(t)= e^{i\omega_r a^\dagger a t}ae^{-i\omega_r a^\dagger a t}$.
By assuming for simplicity that the longitudinal interactions $\gamma(t)$ [and consequently the exchange interaction $J(t)$]  are abruptly turned on at $t=0$, we obtain the infidelity 
  \begin{align}
\label{eq:infid-transv_RWA}
1-\mathcal{F}_T=  \frac{\eta^2\omega_r T_g}{2\pi (1- X^2)^2} \left[ f(X)+ \bar{n} f(X)^* \right]\ , \ \text{with} \ f(X)=2+X \left(i \pi  \left(X^2-1\right)+4 i e^{\frac{i \pi  X}{2}}+2 X\right)  \ ,
\end{align}
and $X= 2(\omega_r-\omega_B) T_g/\pi $. We also introduce the average number of photons  $\bar{n}=|\beta|^2$  in the cavity at the beginning of the two-qubit gate.
At large values of $\omega_r T_g \approx \pi X/2$, we estimate $|1-\mathcal{F}_T|\approx 0.8  (1+\bar{n}) \eta^2$. Using $\eta=1\%$, corresponding to $3.6^\circ$ degrees of misalignment of the magnetic field from the direction of the spin-orbit vector $\textbf{e}_{so}$ and resulting in a realistic suppression of the Rabi frequency of  two orders of magnitude~\cite{AK_prep}, we obtain an infidelity below the surface code threshold  $|1-\mathcal{F}_T|\lesssim 7\times 10^{-4}$ for coherent states with up to $\bar{n}=8$ photons.

\end{document}